\documentclass[twocolumn, trackchanges]{aa}

\newcommand{\degree}{^{\mathrm{o}}}

\newcommand{\micron}{$\rm \mu m$}
\newcommand{\Mjup}{$\rm M_{Jup}$}
\newcommand{\Rjup}{$\rm R_{Jup}$}
\newcommand{\Msun}{$\rm M_{\odot}$}

\newcommand{\Mearth}{$\rm M_{\oplus}$}
\newcommand{\mJyb}{$\ \rm mJy\ beam^{-1}\ $}
\graphicspath{{./}{figs/}}
\usepackage{color} 
\usepackage{chngcntr}
\usepackage[hang,multiple]{footmisc} % for multiple footnotes
\usepackage{hyperref}

\hypersetup{backref=true,pagebackref=true,hyperindex=true,breaklinks=true,colorlinks=true,urlcolor=blue,linkcolor=blue,citecolor=blue,pagecolor=red,bookmarks=true,bookmarksopen=true}

\usepackage{graphicx}
\usepackage{txfonts}

\begin{document} 

   \title{Highly structured disk around the planet host PDS\,70 revealed by high-angular resolution observations with ALMA}
   \titlerunning{PDS 70 ALMA observations}
   \authorrunning{M. Keppler et al.}

   \author{
    M. Keppler\inst{1}
    \and{R. Teague}\inst{2}
    \and{J. Bae}\inst{3}
    \and{M. Benisty}\inst{4,5,6}
    \and{T. Henning}\inst{1}
    \and{R. van Boekel}\inst{1}
    \and{E. Chapillon}\inst{7,8}
    \and{P. Pinilla}\inst{9}
    \and{J. P. Williams}\inst{10}
    \and{G. H.-M. Bertrang}\inst{1}
    \and{S. Facchini} \inst{11}
    \and{M. Flock}\inst{1}
    \and{Ch. Ginski}\inst{12,13}
    \and{A. Juhasz}\inst{14}
    \and{H. Klahr}\inst{1}
    \and{Y. Liu}\inst{15,1}
    \and{A. M\"{u}ller}\inst{1}
    \and{L. M. P\'erez}\inst{4}
    \and{A. Pohl}\inst{1}
    \and{G. Rosotti}\inst{13}
    \and{M. Samland}\inst{1}
    \and{D. Semenov}\inst{1,16}
    }

    \institute{
    Max Planck Institute for Astronomy, K\"{o}nigstuhl 17, 69117, Heidelberg, Germany
    \and Department of Astronomy, University of Michigan, 311 West Hall, 1085 S. University Avenue, Ann Arbor, MI 48109, USA
    \and{Department of Terrestrial Magnetism, Carnegie Institution for Science, 5241 Broad Branch Road, NW, Washington, DC 20015, USA}
    \and{Departamento de Astronom\'ia, Universidad de Chile, Camino El Observatorio 1515, Las Condes, Santiago, Chile}
    \and{Unidad Mixta Internacional Franco-Chilena de Astronom\'{i}a, CNRS, UMI 3386}
    \and{Univ. Grenoble Alpes, CNRS, IPAG, 38000 Grenoble, France.}
    \and{Institut de Radioastronomie Millim\'{e}trique (IRAM), 300 rue de la Piscine, 38406 Saint Martin d’H\`{e}res, France} 
    \and{Laboratoire d’astrophysique de Bordeaux, Univ. Bordeaux, CNRS, B18N, allee Geoffroy Saint-Hilaire, 33615 Pessac, France}
    \and{Department of Astronomy/Steward Observatory, The University of Arizona, 933 North Cherry Avenue, Tucson, AZ 85721, USA}
    \and{Institute for Astronomy, University of Hawaii at Manoa, Honolulu, HI 96822, USA}
    \and{European Southern Observatory, Karl-Schwarzschild-Str. 2, 85748 Garching, Germany}
    \and{Anton Pannekoek Institute for Astronomy, University of Amsterdam, Science Park 904, 1098 XH Amsterdam, The Netherlands}
    \and{Leiden Observatory, Leiden University, P.O. Box 9513, 2300 RA Leiden, The Netherlands}
    \and{Institute of Astronomy, University of Cambridge, Madingley Road, Cambridge CB3 OHA, UK}
    \and{Purple Mountain Observatory \& Key Laboratory for Radio Astronomy, Chinese Academy of Sciences, 2 West Beijing Road, Nanjing 210008, China}
    \and{Department of Chemistry, Ludwig Maximilian University,
Butenandtstr. 5-13, 81377 Munich, Germany}
}

   \date{Received ---; accepted ---}

  \abstract
  % context heading (optional)
    {Imaged in the gap of a transition disk and found at a separation of about 195 mas ($\sim$22 au) from its host star at a position angle of about 155$\degree$, PDS\,70\,b is the most robustly detected young planet to date. This system is therefore a unique laboratory for characterizing the properties of young planetary systems at the stage of their formation. }
  % aims heading (mandatory)
    {We aim to trace direct and indirect imprints of PDS\,70\,b on the gas and dust emission of the circumstellar disk in order to study the properties of this $\sim$5 Myr young planetary system.}
  % methods heading (mandatory)
    {We obtained ALMA band 7 observations of PDS\,70 in dust continuum and $\rm ^{12}CO\,(3-2)$ and combined them with archival data. This resulted in an unprecedented angular resolution of about 70\,mas ($\sim$8\,au). }
  % results heading (mandatory)
    {We derive an upper limit on circumplanetary material at the location of PDS\,70\,b of $\sim$0.01~\Mearth \ and find a highly structured circumstellar disk in both dust and gas.
    The outer dust ring peaks at 0.65\arcsec \ (74 au) and reveals a possible second unresolved peak at about 0.53\arcsec \ (60 au). The integrated intensity of CO also shows evidence of a depletion of emission at $\sim$0.2\arcsec \ (23\,au) with a width of $\sim$0.1\arcsec \ (11\,au).
    The gas kinematics show evidence of a deviation from Keplerian rotation inside $\lesssim$0.8\arcsec \ (91 au). This implies a pressure gradient that can account for the location of the dust ring well beyond the location of PDS\,70\,b. Farther in, we detect an inner disk that appears to be connected to the outer disk by a possible bridge feature in the northwest region in both gas and dust.
    We compare the observations to hydrodynamical simulations that include a planet with different masses that cover the estimated mass range that was previously derived from near-infrared photometry ($\sim$5-9 \Mjup). We find that even a planet with a mass of 10\,\Mjup \ may not be sufficient to explain the extent of the wide gap, and an additional low-mass companion may be needed to account for the observed disk morphology. 
}
   {}


   \maketitle
%
%-------------------------------------------------------------------
\section{Introduction} \label{sec:intro}
In recent years, high angular resolution observations of protoplanetary disks have revolutionized our view of disk evolution and showed that small-scale structures such as concentric rings and spiral arms are ubiquitous \citep[e.g.,][]{Andrews+18,FLong+18}, suggesting that planet formation might occur very early in the history of a young stellar system. Although these substructures are often interpreted as direct imprints of planet-disk interactions, it is still challenging to understand and constrain the architectures of planetary systems that are needed to account for them \citep[e.g.,][]{Bae+18}, or to rule out alternative scenarios \citep[e.g., magneto-hydrodynamical instabilities,][]{ruge2016,Flock+17}. In addition, an accurate determination of the properties of young planets (e.g., luminosity and mass at a given age) is needed to constrain the formation mechanisms that are at work \cite[e.g.,][]{Mordasini17}. 

The theory of interactions of embedded planets with their natal environment, the protoplanetary disk, and their relation to the observational signatures have been studied by many authors \citep[e.g., ][]{Paardekooper+04,Jin+16,Dipierro+18,Liu+18,Zhang+18}. 
Currently the most promising methods for understanding the interaction of a young planet with its environment and its further evolution are to detect it through direct imaging \citep[e.g.,][]{keppler18} or through the perturbations that it induces in the velocity field of the field \citep[e.g.,][]{perez2015}.

Current direct-imaging infrared surveys reach detection limits of a few Jupiter masses \citep[e.g.,][]{maire2017,uyama2017}, but are often limited by bright and complex disk features. Numerous claims of companion candidates in disks that show asymmetric features are indeed still debated \citep[e.g., HD100546, HD169142, MWC758, LkCa15; see][]{quanz2015,Currie15,Follette17,rameau2017,biller2014,reggiani2014,ligi2018,Reggiani2018,Kraus+Ireland12,Sallum+15,Mendigutia+18} and require confirmation through additional observations at different filter bands, for example. 

The presence of three different planets in the disk around HD\,163296 was claimed by two teams with a complementary method based on perturbations in the Keplerian velocity field of the disk. \citet{pinte2018} detected a localized (both in space and velocity) deformation of the isovelocity curves in $^{12}$CO transitions that was consistent with the spiral wake induced by a 2\,M$_{\rm{Jup}}$ planet at 260\,au. \citet{Teague18} measured the rotation velocity curves of CO isotopologues as a function of distance to the star and found local pressure gradients consistent with gaps carved by two $\sim$1\,M$_{\rm{Jup}}$ planets at 83\,au and 137\,au. 

Using the Spectro-Polarimetric High-contrast Exoplanet REsearch instrument on the Very Large Telescope (VLT/SPHERE) and complementary datasets covering multiple epochs and various near-infrared (NIR) wavelengths, we recently discovered a companion to the 5.4$\pm$1.0\,Myr old \citep{mueller2018} and 113.4$\pm$0.5\,pc distant \citep{gaia16,gaia18} T Tauri star PDS\,70 \citep{keppler18,mueller2018}. Comparison of the NIR photometry to evolutionary models implies that the companion is in the planetary mass regime \citep[$\sim$5-9  $\mathrm{M_{\rm Jup}}$,][]{keppler18} which is consistent with the mass range inferred from atmospheric modeling \citep[$\sim$2-17 $\mathrm{M_{Jup}}$,][]{mueller2018}. 
PDS\,70\,b is located at a projected separation of about 22 au from the central star, within the large gap of the transition disk of its host, between an inner disk and a well-resolved outer disk \citep{Hashimoto12,Hashimoto15,Long18,keppler18}. Follow-up direct-imaging observations with the Magellan Adaptive Optics telescope (MagAO) in the H$\alpha$ line enabled a 2-3$\sigma$ detection of the companion at two different epochs. The observations imply that it is likely still accreting gas from the disk \citep{Wagner2018}. This object is therefore a unique case of a directly imaged planet that still shapes its natal environment. 

In this paper, we present new ALMA band 7 observations of PDS\,70 obtained in Cycle 5. We combined the data with archival observations presented by \cite{Long18}, thereby obtaining an unprecedented angular resolution of $\sim$0.07\arcsec. In Sect. \ref{sec:observations} we describe the observing setup and data reduction, and Sect. \ref{sec:results} presents our results, which are discussed and compared to hydrodynamical simulations in Sect. \ref{sec:discussion}. 

%--------------------------------------------------------------------

\section{Observations and data reduction}\label{sec:observations}
We obtained ALMA Cycle 5 director discretionary time (DDT) observations (Project ID: 2017.A.00006.S, PI: M. Keppler) of PDS~70 in band 7 on 2, 3 and 6 December 2017 under very good weather conditions (mean precipitable water vapor, pwv, $\leq$ 0.9 mm).  For three of the four spectral windows, the correlator was tuned to a center frequency of 357.2, 355.3, and 344.3 GHz for continuum observations in dual-polarization mode with a bandwidth of 2.0 GHz. The fourth spectral window was centered around the $^{12}$CO(3-2) transition at 345.8 GHz with a bandwidth of 0.938 GHz. 
The quasars J1427-4206, J1337-1257, and J1517-2422 were used as bandpass, phase, and flux calibrators.  The calibration was performed using the Common Astronomy Software Package (CASA), version 5.1.1. The total on-source integration time was 1.9 hours.

In addition to the $^{12}$CO\,J=3-2 line, we detected emission from the HCN (J=4-3; 354.505 GHz), HCO$^{+}$ (J=4-3; 356.734 GHz), and  H$^{13}$CN (J=4-3; 345.340 GHz) lines. In this paper, we focus on the dust continuum and $^{12}$CO emission, however.

Because the extended antenna configuration filters out the largest spatial scales in the disk, we made use of the archival Cycle 3 data taken in a similar spectral setup and presented by \cite{Long18} to recover the short baselines. Details regarding the observing strategy and setup are described in \cite{Long18}. 
We transferred both Cycle 3 and Cycle 5 data to {\tt\string CASA v.5.3.0} and subtracted the continuum emission from the line data using the task {\tt\string UVCONTSUB}. We corrected the phase center of the Cycle 3 data for the shift due to the proper motion of the star \citep[(-29.7, -23.8) mas/yr,][]{gaia16,gaia18} with respect to to the Cycle 5 data set. 
We then combined the two data sets and shifted the phase center by an amount of (0.509\arcsec, 0.490\arcsec),  which was found to be the center of the disk by fitting a two-dimensional Gaussian to the continuum Cycle 5 emission using the {\tt\string UVMODELFIT} tool.
We finally used the task {\tt\string TCLEAN} for imaging, applying Briggs weighting with a robust parameter of 0.5. Because self-calibration of both continuum and CO data did not significantly improve the images, we will base our analysis on the non-self-calibrated data. 
The resulting beam size for the dust continuum at a mean frequency of 350.6\,GHz (855$\ \mu$m) is $74\times57$ mas (8.4$\times$6.5 au) with a position angle ($PA$) of 63$\degree$. We measured an rms noise level of 0.026 \mJyb from emission-free regions. For the CO, we obtained a beam size of $76\times61$ mas (8.6$\times$6.9 au) with a $PA$ of $60\degree$ and a channel width of 425 m/s. The noise level per channel is determined to be 1.26 \mJyb.\\

%--------------------------------------------------------------------

\section{Results}\label{sec:results}

\begin{figure*}
\centering
\includegraphics[]{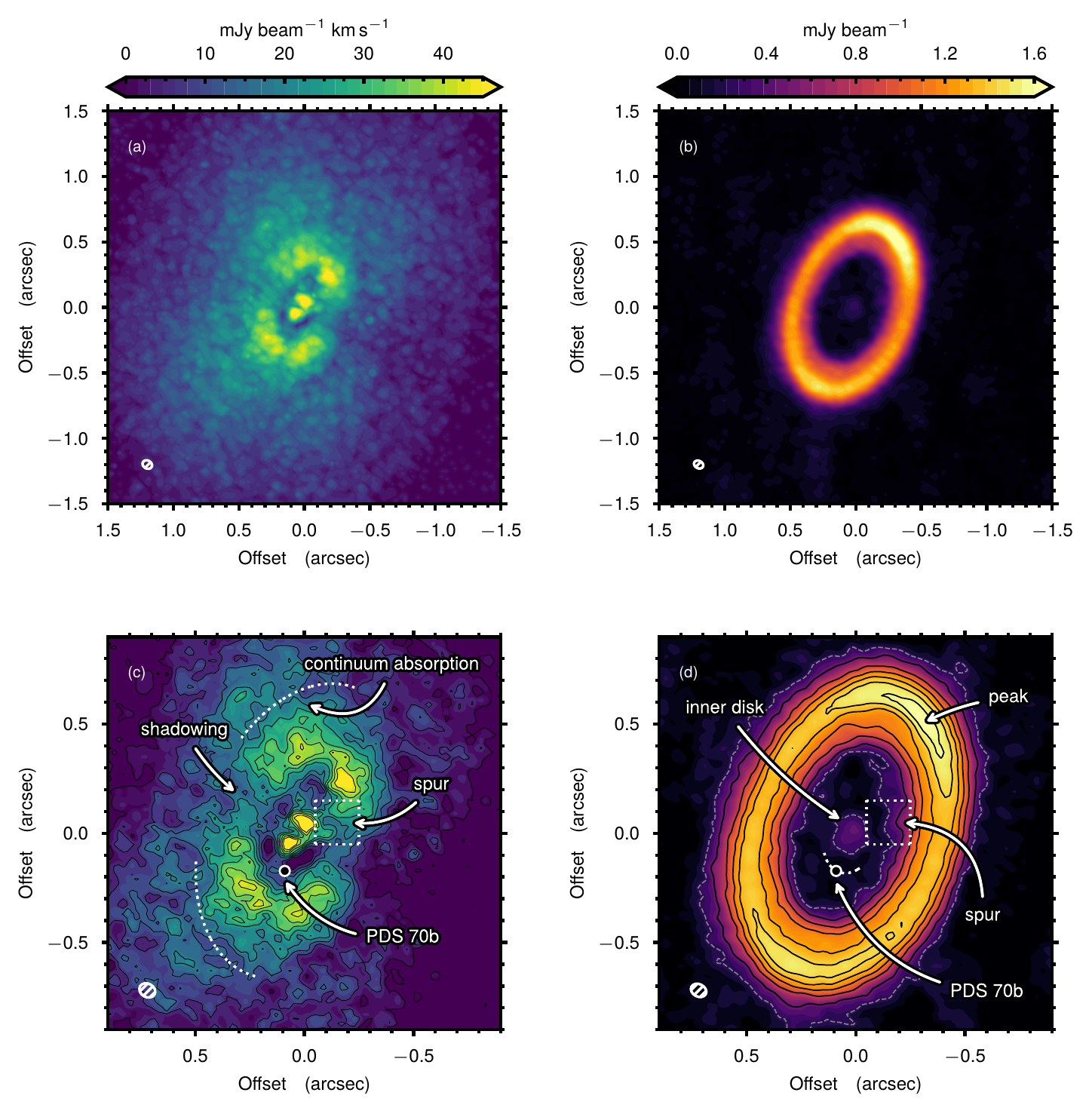}
\caption{Observations of the $^{12}$CO (left column) and the 350.6~GHz continuum (right column). The bottom row provides a closer view of the observations including annotations where the color scaling has been stretched to bring out detail. The contours for the $^{12}$CO are starting at 20\% of the peak value to the peak in steps of 10\%. For the continuum, the gray dashed contour is $5\sigma$, and black contours start at $10\sigma$ and increase in steps of $10\sigma$, where $\sigma = 26~\mathrm{\mu Jy}~{\rm beam}^{-1}$. The synthesized beams are shown in the bottom left corner of each panel. \label{fig:observations}}
\end{figure*}

\subsection{855 $\mu$m dust continuum }
% image properties
Figure \ref{fig:observations} (right column) shows the continuum image at 350.6\,GHz (855\,$\mu$m). 
The disk is detected at a high signal-to-noise ratio (S/N; $\sim$65 at the peak). The integrated flux density of the disk inside $1.3\arcsec$ after applying 2$\sigma$ clipping is 230$\pm$23 mJy, where the error bar is dominated by the $\sim$10$\%$ uncertainty of the absolute amplitude calibration of ALMA in Band 7\footnote{see \url{https://almascience.nrao.edu/documents-and-tools/cycle6/alma-proposers-guide}}. This is consistent with the value found by \cite{Long18}.
The dust continuum shows evidence of a large cavity, a dust ring with a brightness distribution that is slightly asymmetric in both radial and azimuthal direction, an inner disk, as well as a possible bridge feature, all of which we describe in the following paragraphs. 

% disk parameters
By fitting a two-dimensional Gaussian to the two data sets using the task {\tt\string UVMODELFIT}, we find a disk inclination of 51.7$\pm$0.1$\degree$ and 52.1$\pm$0.1$\degree$ and a $PA$ of 156.7$\pm$0.1$\degree$ and 159.7$\pm$0.1$\degree$, for the Cycle 5 and Cycle 3 data sets, respectively. 
We verified the inclination using only short baselines ($<$150\,k$\lambda$, which correspond to the location of the null in the real part of the visibilities, see Fig. \ref{fig:MCMC_dust}) for the Gaussian fit, which ensures that the cavity is not resolved, as well as by using a disk model. These efforts yielded similarly good fits in all cases; the values for the inclination were consistently within 3$\degree$. We note, however, that all these models assume axial symmetry and therefore none of them reproduces the real morphology of the disk. 
Considering the complexity of the continuum emission that appears to be highly structured, such simple modeling appears limited. We adopt a final value of 51.7$\degree$ because this corresponds to the model with the fewest assumptions. 

% disk radial profile
\subsubsection{Disk radial and azimuthal morphology}\label{sec:disk_morphology}
\begin{figure}
\centering
\includegraphics[width=0.5\textwidth]{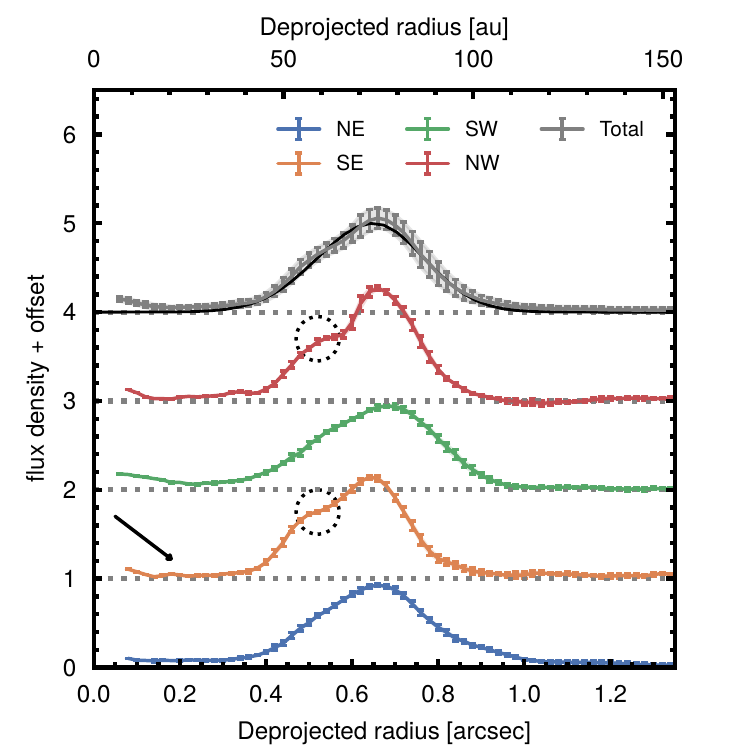}
\caption{Radial profiles of the deprojected dust continuum image along the semi-major (red, orange) and the semi-minor (green, blue) axes, as well as averaged over the entire azimuth (gray). The black line in the uppermost plot corresponds to the best-fit model of the radial profile found in Sect. \ref{sec:disk_morphology}. The deprojection assumes that the continuum is geometrically flat. Radial samples are taken every $\sim$1/4 beam (20 mas), and the cuts along the minor and major axes are azimuthally averaged in a cone of $\pm$10$\degree$ around the corresponding axes. The black arrow highlights a bump in the profile close to the location of PDS\,70\,b, and the dotted circles mark the location of the second peak.}
\label{fig:continuum_profiles}
\end{figure}

Figure \ref{fig:continuum_profiles} (uppermost, gray line) shows the azimuthally averaged and deprojected radial profile of the dust continuum, which clearly reveals a large gap and a ring component. The emission strongly decreases inside the ring, where the flux is reduced by more than 90$\%$. 

The radial profile of the ring is asymmetric, which is best seen in the cuts along the major and minor axes (Fig. \ref{fig:continuum_profiles}, colored lines). The inner edge of the continuum ring reveals a second peak located at a deprojected distance of about 0.53\arcsec \ (60 au). The feature is most pronounced along the major axes, which can be explained by the projection effect as well as by the beam, whose major axis is oriented roughly along the minor axis of the disk. Observations at even higher angular resolution are required to quantify this structure in greater detail. 

To quantify the radial brightness distribution of the dust ring, we used the same approach as \cite{2018ApJ...859...32P}. We first deprojected the data assuming an inclination of 51.7$\degree$, and fit the real part of the deprojected visibilities with a radially asymmetric Gaussian ring using a Markov chain Monte Carlo (MCMC) method using \textit{emcee} \citep{Foreman-Mackey+13}. 
The best-fit model has a peak radial position of 73.7$\pm$0.1 au, and an inner and outer width of 14.8$\pm$0.1 au and 13.4$\pm$0.1 au. The ring is therefore radially resolved by our observations. 
The best fit model is overplotted in Fig. \ref{fig:continuum_profiles} (black line) and is shown in Fig. \ref{fig:MCMC_dust}. 

We confirm the azimuthal brightness enhancement of the ring that was reported by \cite{Long18} on the northwest side of the disk, which peaks at a PA of $\sim$327$^{\circ}$ and is roughly 13$\%$ brighter than the opposite disk side.\footnote{Value found by comparing the peak pixel value of the northwest side with the peak pixel value of the southeast side.}
If the dust is optically thin, the asymmetry could trace the presence of an overdensity. As we argue below, the dust is likely almost optically thick. The brightness enhancement is therefore likely a combination of differences in mass density and temperature. Observations at longer wavelengths are required to break the degeneracies of temperature and density effects and to conclude on the origin of the azimuthal brightness asymmetry. 

% feature: inner disk 
\subsubsection{Inner disk} Our image also confirms the detection of a compact signal toward the location of the star, which has been detected and attributed to be a possible inner disk component by \cite{Long18} the existence of which is consistent with the NIR excess detected in the spectral energy distribution (SED). 
Our observations marginally resolve the emission inside the innermost $\sim$80\,mas (9 au) at a 5$\sigma$ level.
Observations at longer wavelengths will enable us to establish the spectral index of this central emission, which is required to exclude the possible contribution from free-free emission. 

\subsubsection{Possible bridge feature} We detect a spur that projects from the dust ring into the gap in the direction of the inner disk at a $PA$ of about 285$\degree$ (referred to as `spur' in Fig. \ref{fig:observations} and best seen in panel (d)). This signal is even more clearly detected in the DDT data alone, which have a slightly higher resolution (71$\times$56 mas, see Fig. \ref{fig:DDT_dust}). It is possible that the signal forms a bridge feature that connects the outer and inner disks. Whereas the spur is detected at high confidence ($>5\sigma$), the continuous connection to the inner disk in the dust continuum remains to be confirmed with deeper observations.
Interestingly, this feature is cospatial with an extended feature found in scattered light \citep[][see Fig.\ref{fig:DDT_dust}]{keppler18,mueller2018}. Furthermore, the CO shows evidence of a feature at that same location that seems indeed to connect the outer and inner disk (see Sect. \ref{sec:12co}). 

\subsubsection{Upper limits on CPD dust mass}
Figure \ref{fig:continuum_profiles} shows that the radial profile along the southeast semi-major axis presents a marginally (S/N ~$\sim$3) enhanced signal at $\sim$0.2\arcsec. This roughly corresponds to the expected location of PDS\,70\,b. We note, however, that flux density variations of similar amplitude are present at several other position angles as well, and the persistence of this signal is therefore to be tested with deeper observations.

Circumplanetary disks (CPD) are expected to have outer radii $R_{\rm out}$ of a fraction ($\sim$30-70$\%$) of the Hill radius $R_{\rm H}$ \citep[e.g., ][]{Quillen+98,DANgelo+03,Ayliffe+09,Szulagyi14}, where $R_{\rm H}=a_{\rm P}\  (M_{\rm P}/3M_{\star})^{1/3}$ and $a_{\rm P}$ is distance of the planet to the star. For a 5 $M_{\rm Jup}$ companion at 22 au, this corresponds to $\sim$0.8-1.9\,au, and the disk is therefore expected to be unresolved. Our measured noise level of 0.026 \mJyb translates into a 5$\sigma$ upper limit on the flux density of an unresolved CPD around PDS\,70\,b of 0.130 \mJyb. 

We compared this value to the theoretically expected emission from a CPD in order to derive an upper limit on the dust mass. For this aim, we followed the approach presented by \cite{Isella+14}, where the dust temperature $T_d$ in the CPD at a given radius $r$ from the planet is described as
\begin{equation}\label{equ:CPD_temp}
    T_d^4(r) = T_{irr,\star}^4 (a_{\rm P}) + T_{irr,p}^4 (r) + T_{acc}^4 (r),
\end{equation}
where $T_{irr,\star}$ is the temperature of the surrounding circumstellar disk heated by the central star at the distance of the planet to the star, $T_{irr,p}$ is the temperature due to the heating by the planet itself, and $T_{acc}$ denotes the contribution from viscous accretion within the CPD.

For $T_{irr,\star}$ we adopted a value of 19 K at a distance of 22 au from the star, which is estimated from our radiative transfer models \citep{keppler18}. 
The irradiation by the planet, $T_{irr,p}$, can be estimated \citep[assuming a CPD aspect ratio of 0.1; ][]{zhu18} as

\begin{equation}\label{equ:irr}
T_{irr,p}(r) = \left( \frac{L_p}{\sigma_{SB} 40 \pi r^2} \right) ^{1/4},
\end{equation}

 where we used $L_{p}\sim$1.5$\times10^{-4} L_{\odot}$ as the luminosity of PDS\,70\,b \citep{mueller2018}.
 Finally, the heating due to accreting material is given by
 \begin{equation}\label{equ:acc}
 T_{acc}^4(r) = \frac{3 G M_p \dot{M}_{acc}}{8 \pi \sigma_{SB} r^3} \left[ 1 - \left( \frac{r_p}{r}\right)^{1/2} \right],
 \end{equation}
 
 where $\dot{M}_{acc}$ is the mass accretion rate onto the planet and $r_p$ is the planetary radius. Following \cite{Wagner2018}, we assumed $\dot{M}_{acc}\,\sim\,10^{-8}\,\rm M_{\rm Jup}\,\rm yr^{-1}$ and $r_P\,\sim$\,3\,\Rjup \ \citep{mueller2018}.
 
 As in \cite{Isella+14}, we assumed a power-law surface density $\Sigma(r) = C\times\ r^{-3/4}$, where C is the normalization constant for the total CPD dust mass $M_d = \int_{r_{in}}^{r_{out}} \Sigma(r) 2\pi r dr $. We therefore computed the expected millimeter flux $F_d$ for a given $M_d$ by integrating the flux density contribution from each radius over the entire CPD:
\begin{equation}
 F_{d} = \frac{2 \pi \mathrm{cos}i}{d^2} \int_{r_{in}}^{r_{out}} \left(1 - \mathrm{exp}\left[- \frac{\Sigma(r) \kappa}{\mathrm{cos}i} \right] \right) \times B_{\nu}(T_d(r)) r dr.
\end{equation}
 Here, $\kappa$ denotes the dust opacity, which we assumed to be 3.5 $\rm cm^2 g^{-1}$ at 855\,\micron, linearly scaled from \cite{Andrews+12}, $B_{\nu}$ is the Planck function evaluated at $T_d$, and $i$ is the CPD inclination, which we assumed to be equal to the inclination of the circumstellar disk (51.7$\degree$). \\

\begin{figure}[hbt]
\includegraphics[width=0.5\textwidth]{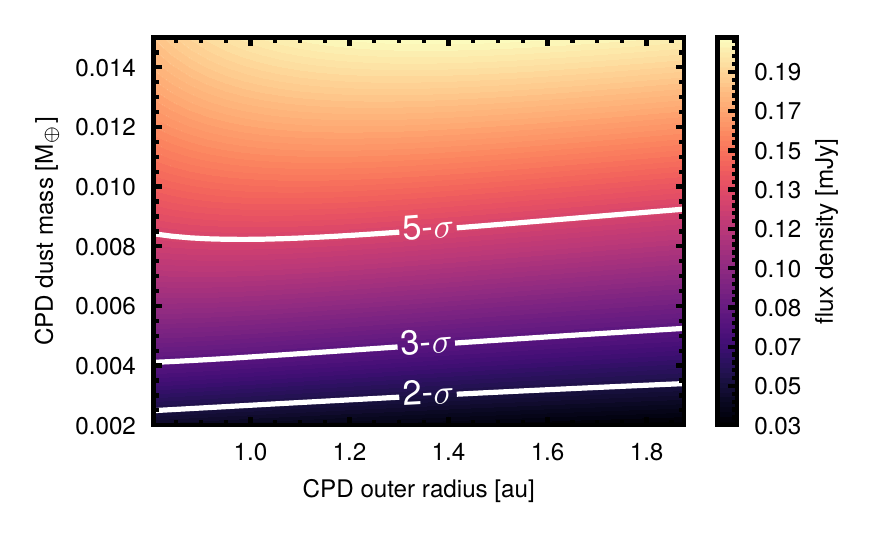}
\caption{Theoretically expected flux densities from a CPD around a 5 \Mjup \ planet at the location of PDS\,70\,b with different dust masses and outer disk radii, following the prescription from \cite{Isella+14}. The contours mark the 2, 3, and 5$\sigma$ detection limits from the observations. }
\label{fig:CPD_limits}
\end{figure}

We computed the expected flux densities for different CPD dust masses considering outer CPD radii of 0.3-0.7 $r_H$ and assuming that the CPD touches the planetary surface (e.g., $r_{in}=r_p$, but note that regions in which the temperature exceeds the sublimation temperature of silicates ($\sim$1500 K) were taken out of the integral). The result was compared to our noise level of 0.026\mJyb and is shown in Fig. \ref{fig:CPD_limits}.
With the given choice of parameters, we find a 5$\sigma$ upper dust mass limit of $\sim$0.01\,\Mearth \ ($\sim$0.8 lunar masses). This value is roughly independent of the outer CPD radius, which means that the emission is likely optically thin. As shown in Appendix \ref{app:CPD}, this detection limit holds for the entire estimated mass range of PDS\,70\,b. 

\subsection{$^{12}$CO $J = 3-2$}
\label{sec:12co}

\begin{figure*}[hbt]
\includegraphics[width=\textwidth]{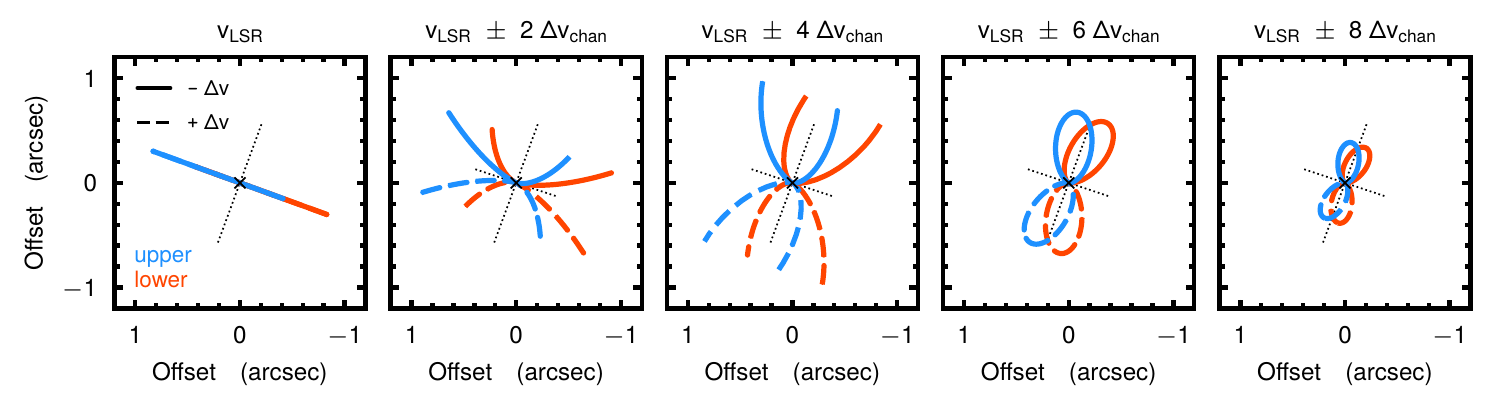}
\caption{Iso-velocity contours for the upper (blue) and lower (red) sides of the disk at different velocities with a flared emission surface. Along the major and minor axes, shown by the black dotted lines, the iso-velocity contours overlap, as in the leftmost and rightmost panels, and thus only emission from the upper side of the disk is visible. Conversely, in inter-axis regions, the iso-velocity contours are spatially separated, as in the central panels, so that emission from both sides of the disk reaches the observer. Based on Fig.~4 from \citet{Rosenfeld+13}.}
\label{fig:isovelocity}
\end{figure*}

\begin{figure*}[hbt]
\centering
\includegraphics[]{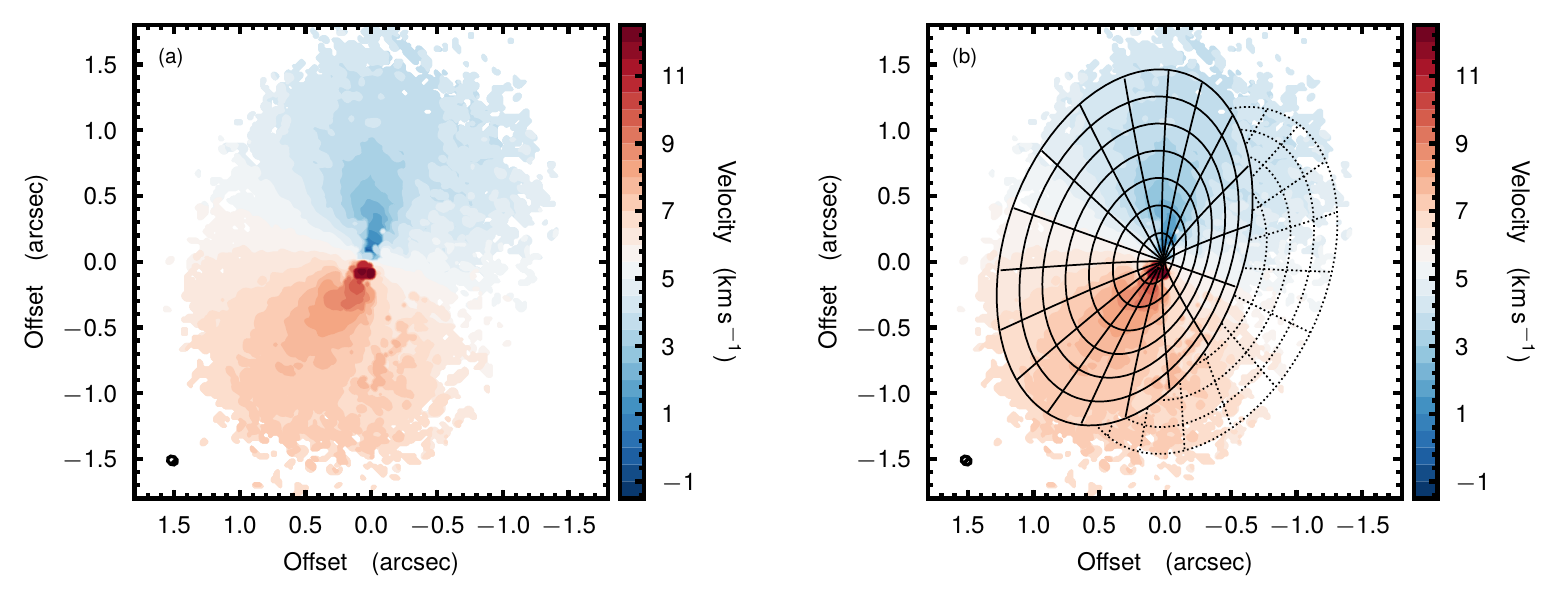}
\caption{Rotation profile of the $^{12}$CO emission, \emph{left}, using the method presented in \citet{Teague_Foreman-Mackey_2018}, with the best-fit surface overlaid, \emph{right}. The solid lines show the top surface, and the dotted lines show the far surface. \label{fig:ninth_moment}}
\end{figure*}

Figures~\ref{fig:observations} (a) and (c) show the $^{12}$CO $J=3-2$ integrated intensity (zeroth-moment) map; panel (c) includes annotations of the main features. The asymmetry with respect to the disk major axis is clear. This is due to the significantly elevated $\tau \sim 1$ surface of the $^{12}$CO, which is typically assumed to trace disk layers where $z \, / r \sim 0.25$ \citep{Rosenfeld+13}. In addition, several other features are visible, including two gaps (a prominent gap at $\sim 0.2\arcsec$ and a faint gap at $\sim 0.6\arcsec$), a bridge-like feature similar to the one observed in the continuum, and apparent shadowing along the major and minor axes that has previously been reported by \citet{Long18}.

Toward the center of the image, the inner disk component is clearly detected, extending out to approximately 15 au, which is consistent with estimates from scattered light \citep{keppler18}. For disks shaped by planets, a bright gaseous inner disk (implying a gas gap rather than a cavity) is in agreement with the predictions from hydrodynamical models, even for the cases where the planet mass is as high as 10 \Mjup \  \citep{Facchini+18}. 

At about the same location as the spur found in the continuum, the zeroth-moment map shows evidence of an extended signal that connects the inner disk and the outer ring in the northwest region. This signal may be connected to the extended feature detected in the NIR \citep[][ and Fig. \ref{fig:DDT_dust}, right panel, of this paper]{keppler18,mueller2018}, and might also be related to the features seen in CO and $\rm HCO^{+}$ by \cite{Long18} at similar locations. 
If this feature indeed connects the outer and inner disks, it may be tracing gas flow through the gap from the outer to the inner disk \citep[e.g., ][]{Tang17,Casassus15,Price+18}. This hypothesis could be confirmed through the detection of localised velocity changes in the given region, which we do not detect with our spectral resolution, however. The nature of this feature therefore needs to be tested with observations at higher spectral and angular resolution. 

The inner gap at $\sim$0.2$\arcsec$ is likely due to a gap opened by PDS\,70\,b and is discussed further in Section~\ref{sec:modelling}. The outer gap at $\sim$0.6\arcsec \ can be explained by continuum absorption of the bottom side of the disk: as shown in Fig. \ref{fig:isovelocity}, the contours of equal projected velocity at the top and bottom sides of the disk in regions between the disk major and minor axes are spatially offset. While emission from the bottom side travels through the midplane toward the observer, it is absorbed by the dust, which reduces the integrated flux at that location \citep[e.g.,][]{Isella+18}. As emission from the bottom side of the disk is almost entirely absorbed, we conclude that the dust ring is likely optically thick at $\nu = 345$~GHz, a result which has found at millimeter wavelengths for other disks as well \citep[e.g.,][]{Pinilla+17}.

Along the disk major and minor axes, on the other hand, the iso-velocity contours do overlap. Because the $^{12}$CO is optically thick, emission from the bottom side of the disk is self-absorbed and only the top side is visible. This causes the apparent shadowing along the major and minor axes of the disk \citep[and the shadowing observed in the HCO$^+$ data presented by][]{Long18}. A more elevated emission layer results in a larger azimuthal variance, because the two sides become more spatially resolved. The difference between the value along an inter-axis region and along an axis will peak at roughly a factor of two, a feature that is commonly seen in the integrated intensity maps of high spatial resolution observations of $^{12}$CO \citep[e.g.,][]{Rosenfeld+13}.

\subsubsection{Deriving a $^{12}$CO emission surface}

Because the $^{12}$CO emission comes from an elevated layer above the midplane, we needed to deproject the data in order to precisely analyze the emission and velocity structure as a function of the radius.  For this aim, we wished to derive constraints on the emission height of the $^{12}$CO. Following \citet{Teague+2018b}, we generated a map of the rotation velocity using the method presented in \citet{Teague_Foreman-Mackey_2018}\footnote{using \texttt{bettermoments} \citep{bettermoments}}, which is robust against confusion from the near and far sides of the disk\footnote{Carrying out this modeling approach on an intensity-weighted average velocity map (first-moment map), we find a much flatter disk due to the averaging of the upper and lower sides of the disk.}. We then fit a Keplerian rotation pattern to the data, including a flared emission surface parameterized as $z(r) = z_0 \times (r \,/ \, 1\arcsec)^{\varphi}$, and fixed the inclination at $i = 51.7\degr$ to break the degeneracy with the stellar mass. 
We note that our modeling of the surface height is limited to a generic model of a flared surface because the resolution of our data is limited. To perform more detailed modeling of the emission surface under consideration of spatial variations of the underlying gas density structure, a higher resolution is required.
Our modeling results in a tight constraint on the emission surface of

\begin{equation}
    z(r) [\arcsec] = \left(0.33 \pm 0.01\right) \times \left( \frac{r}{1\arcsec} \right)^{0.76 \pm 0.01},
\end{equation}

\noindent with the additional parameters of $M_{\rm star} = 0.875 \pm 0.03~M_{\rm sun}$, ${\rm PA} = 160.4\degr \pm 0.1\degr$, and $v_{\rm LSR} = 5505 \pm 2~{\rm m\,s^{-1}}$. These uncertainties describe the 16th to 84th percentile range of the posterior distributions for each parameter which are symmetric about the median. We note that these uncertainties correspond to the statistical uncertainties and do not take into account the systematic uncertainties that may be significantly larger. Figure~\ref{fig:ninth_moment} shows the best-fit emission surface overlaid on the rotation map.

\begin{figure}
\includegraphics[width=\columnwidth]{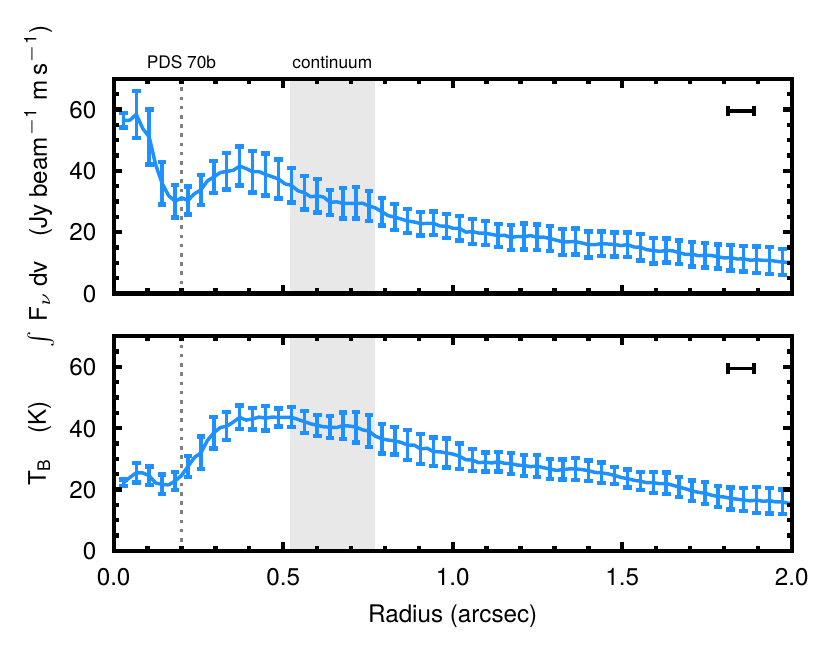}
\caption{Radial profiles of the $^{12}$CO integrated intensity, \emph{top}, and brightness temperature, \emph{bottom}. Radial samples are taken every 1/4 beam and the error bar shows the standard deviation in the azimuthal bin. The vertical dotted line shows the orbit of PDS~70b, while the gray shaded region shows the extent of the continuum ring. The beam size is shown in the top right corner of each panel. \label{fig:zeroth_moment_profiles}}
\end{figure}

Using this emission surface, the data were deprojected into bins of constant radius and were azimuthally averaged with the resulting integrated intensity profiles shown in Fig.~\ref{fig:zeroth_moment_profiles}. The radial profile of the integrated flux density in the top panel shows a clear gap at $0.2\arcsec$ ($\sim$23~au), consistent with the orbit of PDS~70b \citep{keppler18, mueller2018} and a gap width of $\sim0.1\arcsec$. Because of the very high optical depth of $^{12}$CO, any visible gap feature requires a significant depletion of gas or considerable change in gas temperature \citep[e.g.,][]{Facchini+18}. 

Using the brightness temperature, $T_{\rm B}$, presented in Fig. \ref{fig:zeroth_moment_profiles} (lower panel) as a proxy of the gas temperature, we infer a drop in the local gas temperature across the gap. This is consistent with a surface density depletion of the gas, which would move the $\tau = 1$ surface of the $^{12}$CO deeper within the disk, closer to the cooler midplane, therefore dropping the temperature. 
One possibility to clearly distinguish the effects of temperature and density on the brightness temperature is to use the CO line width as a tracer for temperature variations \citep{Teague18}, for which higher spectral resolution is required than given by our data, however.

From the integrated flux density profile, we find that the gap extends from about 0.1 to 0.3$\arcsec$ ($\sim$11 to 34 au). It is spatially resolved, and does not seem to extend out to the location of the dust continuum ring, although it is not possible to measure the $^{12}$CO depletion accurately because of its large optical depth. 
This preferential depletion of grains compared to gas within a cavity is a common feature for transition disks \citep{vanderMarel_ea_2015, vanderMarel_ea_2016}.

\subsubsection{$^{12}$CO rotation curve}\label{sect:rotation_curve}

\begin{figure}
    \centering
    \includegraphics[width=\columnwidth]{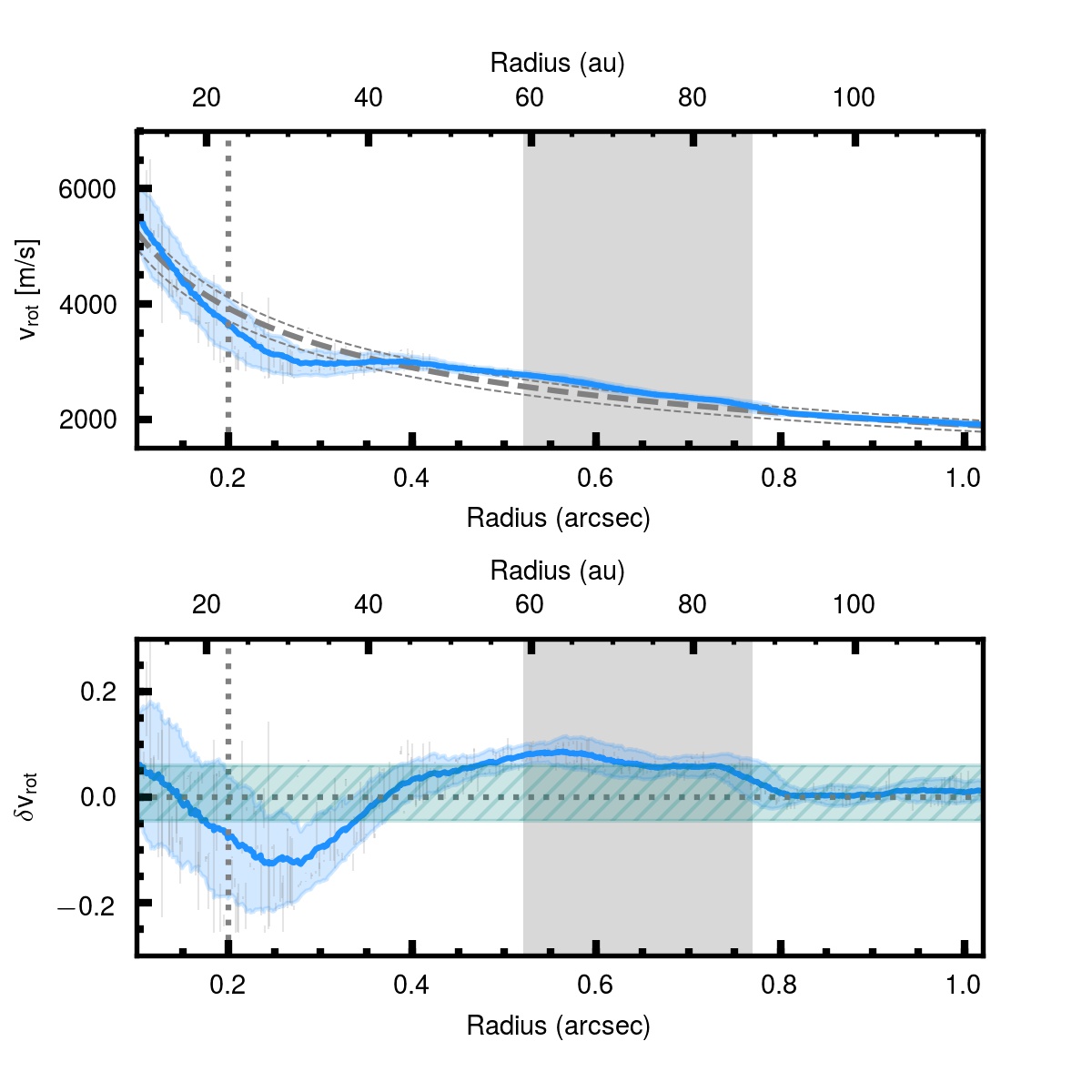}
    \caption{\emph{Top:} Measured rotation curve with $1\sigma$ uncertainties. The blue line and blue shadowed area show the running mean and its standard deviation. The dashed gray lines show the Keplerian rotation curve assuming the best-fit stellar mass (0.88 \Msun, thick) and including the 3$\sigma$ uncertainties on the stellar mass (corresponding to 0.79 and 0.97 \Msun \ respectively, thin) derived from the rotation map fitting. The uncertainties of the stellar mass correspond to the statistical uncertainties and do not include the systematics.  \emph{Bottom:} Relative residuals (blue solid) and uncertainties (blue shaded area) between a smooth Keplerian curve and the inferred rotation curve.  
    The green hatched area highlights the uncertainty of the absolute scaling of $\rm \delta v_{rot}$ inferred by the 3$\sigma$ statistical uncertainties on the stellar mass. In both panels, the gray shaded region shows the extent of the continuum ring. 
    The vertical dotted line shows the orbit of PDS\,70\,b and the shaded vertical gray region traces the location of the continuum emission. 
    }
    \label{fig:rotation_curve}
\end{figure}

Radial gas pressure gradients perturb the gas rotation velocity and are used as tracers for planet-induced perturbations \citep{perez2015,pinte2018,Teague18}. 
Velocity distortions by the planet at the close-in location of 22 au are small, such that their detection in single-channel maps as described by \cite{pinte2018} is hampered by our limited angular and spectral resolution (see also Sect. \ref{sec:comp_model_obs}), and further by the relatively low S/N of the CO emission at the location of the planet (well within the CO-integrated flux density gap). 
To improve the S/N of potential kinematic perturbations we therefore used of an azimuthally averaged rotation curve of the $^{12}$CO data to probe the underlying gas density structure \citep{Teague18}. This is even possible in cases when the line emission is optically thick.
Whereas a negative pressure gradient induces sub-Keplerian rotation, a positive pressure gradient would cause super-Keplerian rotation. 

Following the method described in \citet{Teague+2018b}, we inferred the rotation profile by determining the rotation velocity for each radius which allows for all spectra in an annulus to be shifted back to the same systemic velocity\footnote{A Python implementation of this method, \texttt{eddy} \citep[see][]{eddy}, is publicly available at \url{https://github.com/richteague/eddy}.}. We ran ten different realizations of this, randomizing the pixels taken from each annulus (making sure they are separated by at least one FWHM of the beam), and randomizing the radial locations of the annuli while maintaining a radial bin width of a quarter beam width. The resulting rotation curve and the residual relative to the best-fit Keplerian profile are plotted in Fig.~\ref{fig:rotation_curve}. 

The absolute scale of the deviation from Keplerian rotation depends on the reference Keplerian velocity and therefore on the assumed stellar mass. The systematic uncertainties on the dynamical determination of the stellar mass as well as the parameterization of the surface together with the fact that our fiducial model for the rotation velocity does not take into account the overall pressure gradient in the disk may cause the uncertainty of the absolute scaling to be as large as 10$\%$. Figure~\ref{fig:rotation_curve} (bottom panel) shows the residuals of the rotation curve (blue), where the green hatched area marks the uncertainty of the zero-point of $\mathrm{\delta v_{rot}}$ inferred by the 3$\sigma$ statistical uncertainties of the stellar mass. Within these uncertainties, the peak of the continuum ring ($\sim$0.65\arcsec) lies close to the location where $\mathrm{\delta v_{rot}}$ recovers Keplerian rotation and therefore where pressure reaches its maximum.

A significant deviation of up to $\sim 12\%$ at $\sim 0.2\arcsec$ is observed, which is suggestive of significant changes in the gas pressure at this location, consistent with the structure observed in the rotation map in Fig.~\ref{fig:ninth_moment}.
The rotation curve clearly demonstrates a positive pressure gradient between $\sim$0.4 and 0.8\arcsec, reaching a maximum at about 0.55\arcsec. This implies that the gas density is likely depleted beyond $\sim$0.4\arcsec, and therefore suggests that the gap is in reality larger than what is observed in integrated emission: if the gap were only as wide as the gap in the $^{12}$CO integrated emission, then we would expect the peak residual of the rotation curve to fall at the edge of the gap at $\sim 0.3\arcsec$ \citep[see Fig.~1 in][for example]{Teague18}, but the peak is found closer to $0.55\arcsec$. 
The shape of the residual curve in the inner disk, r $<$ 0.3\arcsec, is dominated by the steep gradients in the intensity profile that are due to both the inner disk and the gap; this makes a direct analysis challenging. A more thorough discussion of this effect and the effect of beam smearing is discussed in Sect. 4.1.2 in the context of hydrodynamical models and in Appendix \ref{app:beam_smearing}.

\subsubsection{Potential point source}

\begin{figure}
\includegraphics[width=\columnwidth]{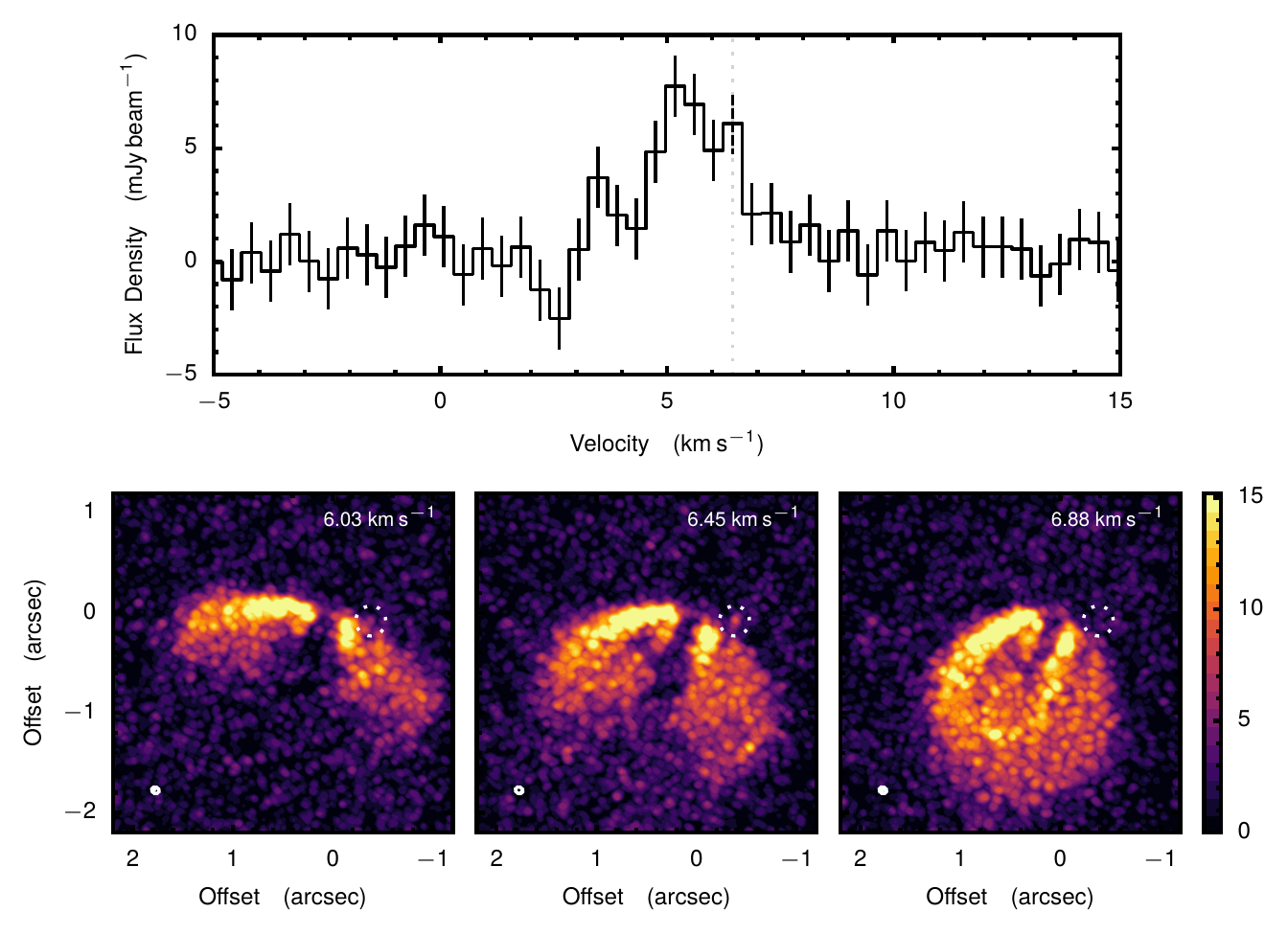}
\caption{Top panel: CO line profile extracted at the location of the point source. Bottom panel: CO channel maps around 6.45 km/s. The white circle indicates the location of the point source.} \label{fig:point_source}
\end{figure}

We tentatively detect a point source in the $^{12}$CO emission maps at a projected separation of $\sim$0.39$\arcsec$ and a PA of $\sim$260$\degree$. 
This corresponds to a deprojected radius of $\sim$71~au, if if comes from the midplane. The peak is detected at a $\sim 6\sigma$ level and is spatially offset from the Keplerian emission pattern. Figure~\ref{fig:point_source} shows the spectrum extracted at the location of the source, and three channel maps showing the offset nature of the emission. The signal appears at a velocity of around 6.45 $\rm km~s^{-1}$, corresponding to a redshift of roughly 1 km/s with respect to the line center of the Keplerian profile. The spectrum also shows a blueshifted peak, whose emission may be biased from the bottom side of the disk, however. Interestingly, if it were located in the midplane, the source would be located well within the dust continuum ring, close to the dip between the main and the tentative second peak detected in the continuum profiles (see Sect. \ref{sec:disk_morphology}). 
Spatially offset emission has been shown to potentially be a signature of a CPD \citep{perez2015}, as the additional rotation of the CPD would shift the emission from the Keplerian pattern. 
If the signal were indeed connected to a forming embedded planet, this might explain the azimuthal gap found in the $\rm HCO^{+}$ emission at a similar location \citep{Long18} because chemical changes due to heating from the planet may locally deplete HCO \citep{Cleeves+15}. 
Additional observations are required to confirm the potential point source. 

%-----------------------------------------------------------------------

\section{Discussion}\label{sec:discussion}

As shown by theoretical studies, the interaction of a massive body with the disk opens a gap in the gas \citep[e.g.,][]{Lin+Papaloizou1986}. The perturbation of the local gas density causes a change in the local pressure gradient, which manifests itself in two ways. First, it generates a pressure bump outside the planetary orbit, trapping large dust particles (while small particles that are well coupled to the gas may still enter the gap). This leads to a spatial segregation of large and small grains \citep[e.g., ][]{Pinilla12}.  
Second, the change in pressure gradient manifests itself in a local deviation from Keplerian rotation the amplitude of which is sensitive to the planet mass \citep{Teague18}. 

Our aim is to investigate the impact of PDS\,70\,b on the observed disk morphology. For this purpose we carried out hydrodynamic and radiative transfer simulations that we present in the next section. 
\\

\subsection{Hydrodynamic and radiative transfer models}
\label{sec:modelling}

\subsubsection{Model setup}
\begin{figure}
\includegraphics[width=\columnwidth]{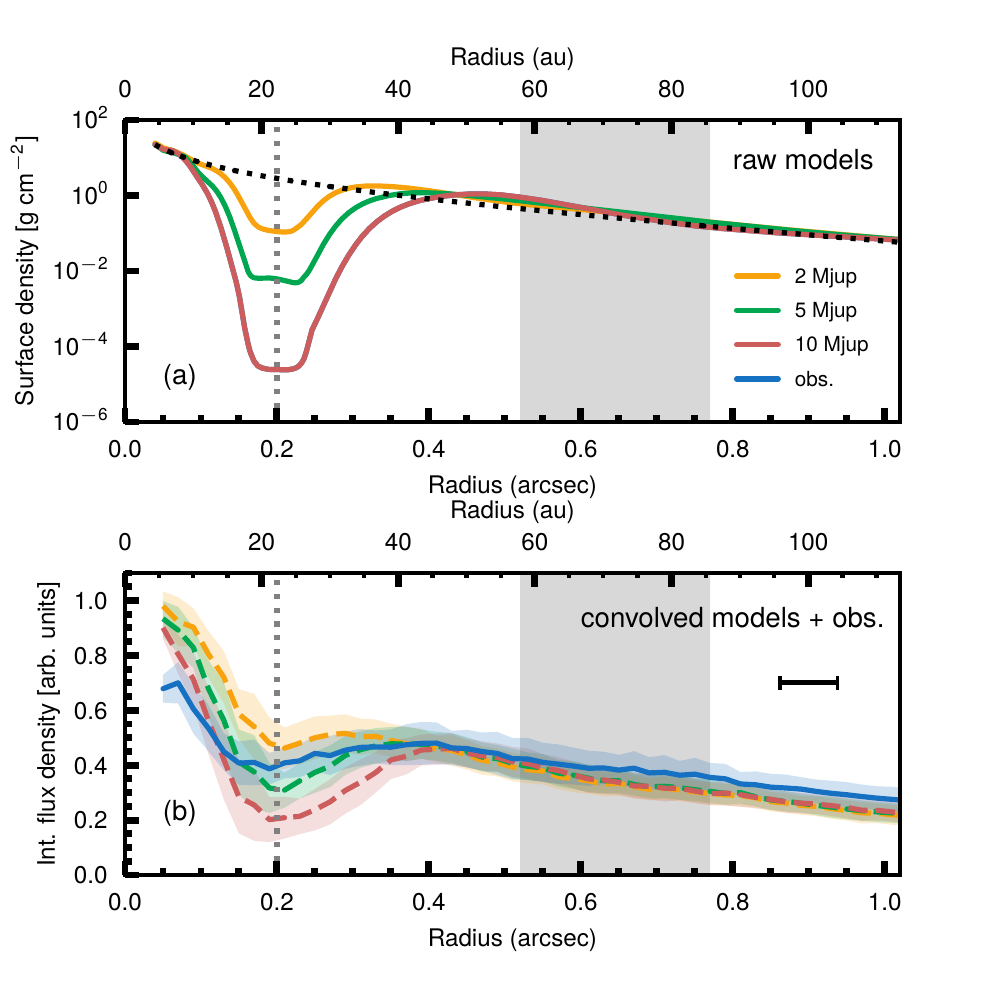}
\caption{
Comparison of hydrodynamical models including a 2~\Mjup\ (yellow), 5~\Mjup\ (green), and 10~\Mjup\ (red) planet located at 0.2\arcsec \ with the observations (blue).
(\textbf{a}): Azimuthally averaged surface density profiles of hydrodynamical simulations. The dotted line corresponds to the initial unperturbed surface density profile.
(\textbf{b}): Integrated azimuthally averaged CO flux density of observations and ALMA-simulated models, after applying 2$\sigma$ clipping. 
In each panel, the gray shaded area indicates the extension of the continuum ring, and the vertical dotted line corresponds to the approximate location of PDS\,70\,b. The black bar in the second panel indicates the major axis of the beam (0.076\arcsec). 
\label{fig:fluxdens}}
\end{figure}

To simulate interaction between PDS\,70\,b and the circumstellar disk of PDS~70, we carried out three-dimensional hydrodynamic calculations using FARGO3D \citep{benitez16,masset00}.
We adopted the disk density and aspect ratio profiles used in \citet{keppler18}
\begin{equation}\label{equ:surfdens}
\Sigma_{\rm gas}(R) = \Sigma_c \left( {R \over R_c} \right)^{-1} \exp \left( -{R \over R_c} \right)
\end{equation}
and 
\begin{equation}
{H \over R} = \left({H \over R}\right)_p \times \left( {R \over R_p} \right)^f,
\end{equation}
where $R_c=40$~au, $R_p=22$~au is the distance of PDS\,70\,b assuming a circular orbit, $(H/R)_p = 0.089$, and $f=0.25$.
$\Sigma_c = 2.87~{\rm g~cm}^{-2}$ was chosen such that the total gas mass in the disk was $0.003~M_\odot$, consistent with the model presented in \citet{keppler18}. The surface density profiles are shown in Fig. \ref{fig:fluxdens} (a). We assumed a vertically isothermal disk temperature structure and used an isothermal  equation of state. 

The simulation domain extends from $r = 0.2~R_p$ to $9~R_p$ in the radial direction, from $\pi/2-0.4$ to $\pi/2$ in the meridional direction, and from 0 to $2\pi$ in the azimuthal direction.
We adopted 256 logarithmically spaced grid cells in the radial direction, 48 uniformly spaced grid cells in the meridional direction, and 420 uniformly spaced grid cells in the azimuthal direction.
A disk viscosity of $\alpha=10^{-3}$ was added to the simulations. This value of turbulence is consistent with the level of turbulence constrained for the protoplanetary disks around TW\,Hya \citep{Teague16,Teague18_twhya,Flaherty18} and HD\,163296 \citep{Flaherty15,Flaherty17}.

\begin{figure}
\includegraphics[width=\columnwidth]{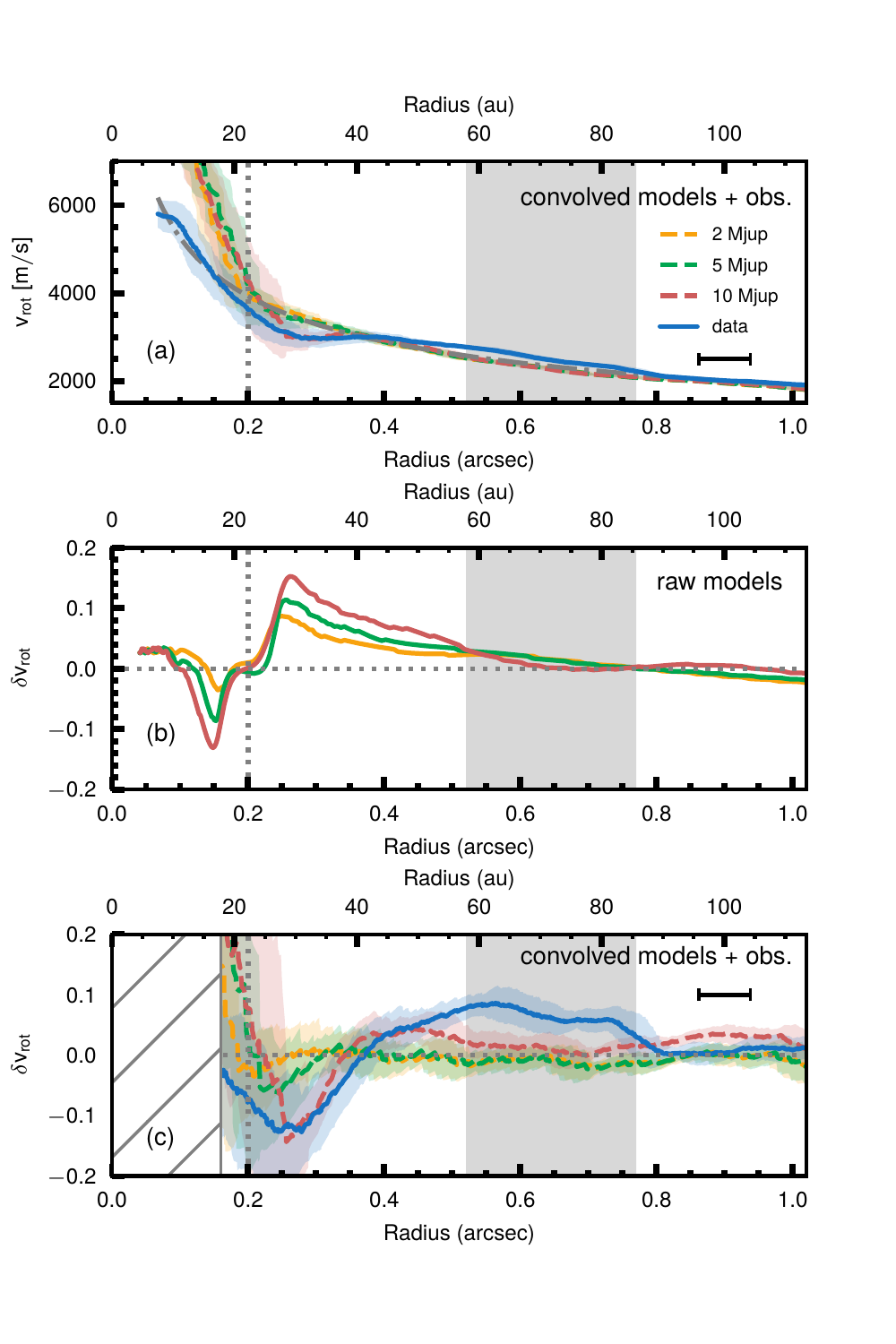}
\caption{
Comparison of hydrodynamical models including a 2~\Mjup\ (yellow), 5~\Mjup\ (green), and 10~\Mjup\ (red) planet located at 0.2\arcsec \ with the observations (blue).
(\textbf{a}): Rotation velocity as a function of deprojected distance. The gray dash-dotted line indicates the unperturbed Keplerian profile around a 0.88 \Msun \ star. 
(\textbf{b}): Deviation from Keplerian rotation of the hydrodynamical simulations at the $\tau$=1 surface. 
(\textbf{c}): Deviation from Keplerian rotation of ALMA-simulated models and observations. The plot shows the running mean and standard deviations. The inner region up to 160 mas is affected by beam confusion effects and is therefore blocked out.
In each panel, the gray shaded area indicates the extension of the continuum ring, and the vertical dotted line corresponds to the approximate location of PDS\,70\,b. The black bar in the first and third panel indicates the major axis of the beam (0.076\arcsec). 
\label{fig:vrot}}
\end{figure}

We tested three planet masses: 2, 5, and 10\,\Mjup, covering the range of potential planet masses proposed by \citet{keppler18}, assuming a 0.85 solar-mass star. The simulations ran for 1000~orbits, after which we find that the gap width and depth reached a quasi-steady state. This is in agreement with other planet-disk interaction simulations from the literature \citep[e.g., ][]{Duffell13,Fung14,Kanagawa15}. The radial profile of the deviations from Keplerian rotation after 1000~orbits is shown in Fig. \ref{fig:vrot} (b).

We generated $^{12}$CO image cubes using the radiative transfer code RADMC3D version 0.41\footnote{\url{http://www.ita.uni-heidelberg.de/~dullemond/software/radmc-3d/}}.
We first computed the thermal structure of the disk by running a thermal Monte Carlo calculation.
To do so, we placed a 0.85 solar-mass star at the center. This star had an effective temperature of 3972~K and a radius of $1.26~\rm R_\odot$  \citep{pecaut16,keppler18}, emitting $10^8$ photon packages.
As in \citet{keppler18}, we considered two grain size distributions whose number density followed a power law as a function of the grain size $a$ with $n(a) \propto a^{-3.5}$: small grains ranged from 0.001 to 0.15 $\mu m$ and large grains ranged from 0.15 to 1000 $\mu m$. 
The relative mass fraction of small to large grains was 1/31, implying that about 3$\%$ of the total dust mass was confined within the small grain population. This is consistent with previous radiative transfer models of PDS\,70 \citep{Dong12,keppler18}.
We assumed that the grains are composed of $70~\%$ astronomical silicates \citep{draine03} and $30~\%$ amorphous carbon grains \citep{zubko96}.
The grain opacity was computed according to the Mie theory using the BHMIE code \citep{bohren83}. 

CO line radiative transfer was done under local thermal equilibrium (LTE) assumptions, assuming a constant $\rm ^{12}CO$ to $\rm H_2$ ratio of $\rm 10^{-4}$ \citep[e.g.,][]{Lacy94,Williams14}. 
A local spatially unresolved microturbulence was added at a constant level of $30~{\rm m~s}^{-1}$. 
This choice is equivalent to $\alpha$ of a few $\times 10^{-3}$.
We simulated the ALMA observations using the {\tt\string SIMOBSERVE} task in CASA version 5.1.2. using the same velocity resolution, synthesized beam, and on-source integration time as were used in the observations.
Thermal noise from the atmosphere and from the antenna receivers was added by setting the {\it thermalnoise} option in the {\it  simobserve} task to {\it tsys-atm}.  
Using the same tools as for the observations, we derived the velocity-integrated flux density, as well as the rotation profiles for each simulation (Figures \ref{fig:fluxdens} and \ref{fig:vrot}).

\begin{figure*}
\centering
\includegraphics[width=0.85\textwidth]{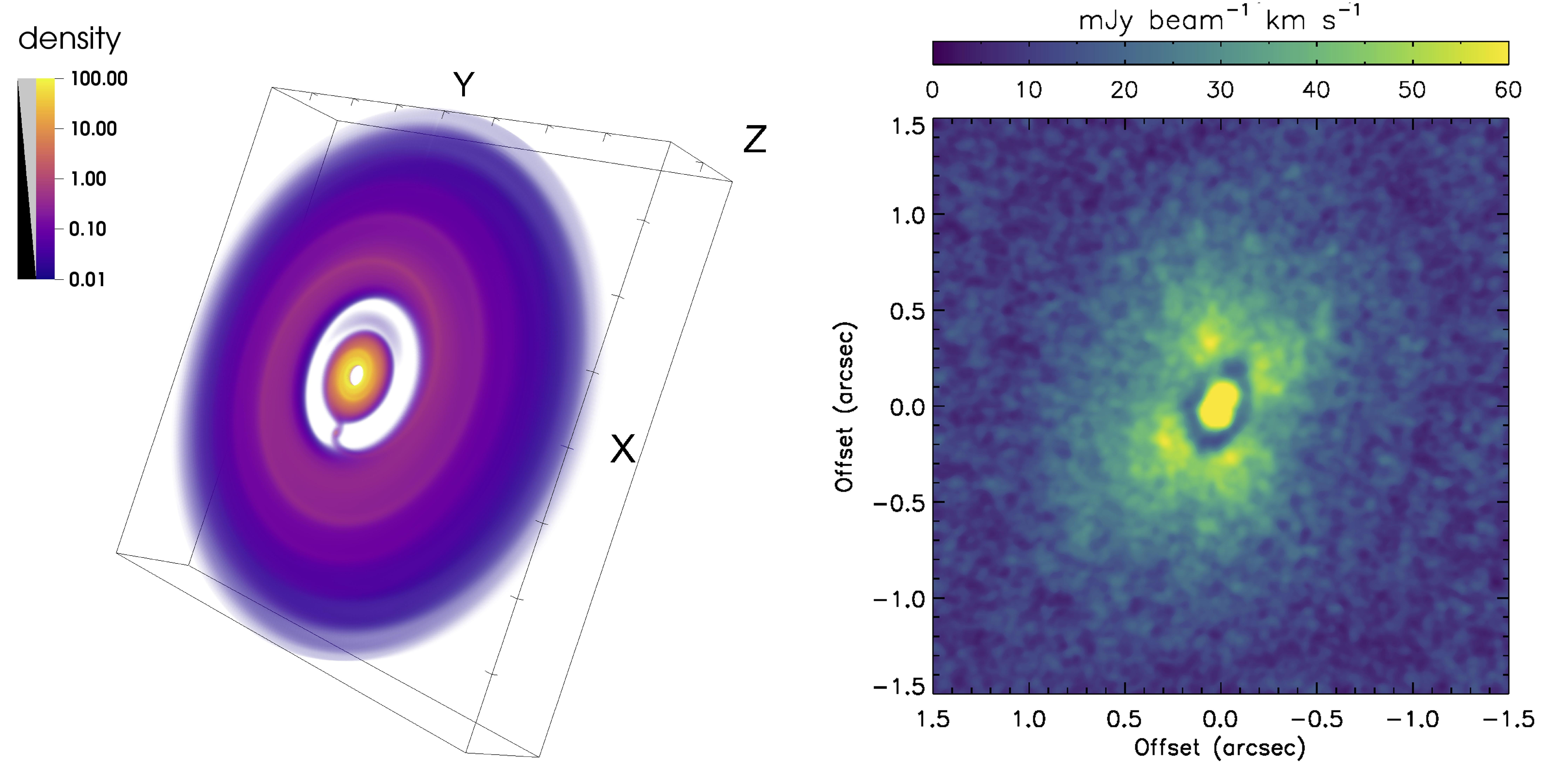}
\caption{\textit{Left:} Three-dimensional volume rendering of the gas density (in a normalized unit with logarithmic scaling) after evolution of 1000 orbits in the inner 100~au of the model disk with a 5 \Mjup \ planet at 22\,au. Ticks on the axes mark every 25~au. \textit{Right:} Simulated $^{12}$CO zeroth-moment map based on the hydrodynamic model presented in the left panel.}
\label{fig:hydromodel}
\end{figure*}

\subsubsection{Comparison with observations}\label{sec:comp_model_obs}

%JB 
The disk density distribution from the hydrodynamic model and a simulated $^{12}$CO zeroth-moment map are presented in Fig.\,\ref{fig:hydromodel}.
The 5~\Mjup \ planet opens a gap around its orbit, which is clearly visible in the simulated zeroth moment map.
We find that velocity kinks associated with the planet-driven spiral arms are present in raw simulated channel maps, similar to what is found in HD~163296 \citep{pinte2018}.
However, the velocity distortions are too small and thus smeared out after convolution with the ALMA beam.

%JB
We compare the radial profiles of the simulated and observed integrated flux densities  in Figure \ref{fig:fluxdens} ($b$). The profiles show evidence of a depletion in integrated flux density at the location of the planet, which is stronger for higher planet masses. 
The width and depth of the depleted flux density in the observations are reasonably well reproduced by a 5~\Mjup \ planet.
We note that the models appear to overestimate the increase in flux density toward the inner disk. 
Because CO is optically thick, this is likely caused by a different temperature structure of the inner disk region, the reason for which could be a different density profile than assumed (e.g., overestimation of the actual density in the inner part of the disk, or a different gap shape), but needs further investigation with higher angular resolution that is able to better resolve these inner regions. 

Figure \ref{fig:vrot} (panel $a$) presents the absolute rotation profiles, and the residual $\delta v_{\rm rot}$ profiles before and after radiative transfer and ALMA simulations are shown in panels $b$ and $c$, respectively. 
We note two points: first, a comparison of the residual model profiles before and after convolution alters the overall shape of the rotation curves. Second, the residual curve of the PDS\,70 disk follows the general shape of the modeled curves, but it differs with respect to the location of the maximum, as well as the velocity gradient toward the inner disk.

The change in the shape of the rotation curve when simulating the observations is due to beam convolution effects in the presence of strong radial gradients in intensity and velocity. This is described in detail in Appendix \ref{app:beam_smearing}. In brief, sharp edges in the flux density profile induce a distortion in the measurement of the rotation curve because the velocities measured within one beam are biased toward those at the highest line intensity. This causes the velocity to be overestimated in the inner region of the gap and underestimated at the outer edge of the gap.  
The resulting rotation curve is of a characteristic shape that is asymmetric with respect to the gap center (see Fig. \ref{fig:beam_smear_analytical}). It shows evidence of strong super-Keplerian rotation in the inner gap region, and a weaker region of sub-Keplerian rotation at the outer gap edge. This effect is now superimposed on the effect of the planet-induced pressure gradient on the rotation profile (sub-/super-Keplerian rotation inside/outside the planetary orbit). This effect can be fully accounted for when performing forward modeling. As Fig. \ref{fig:vrot} ($c$) shows, all convolved model profiles show this characteristic shape, and the amplitudes of their minima and maxima depend on the planetary mass. 

The observed rotation curve of PDS\,70 shows the same characteristic transition from sub-Keplerian to super-Keplerian transition as the models. While we found that the width and depth of the integrated flux density profiles seem consistent with the effect of a 5~\Mjup \ planet, we find that the radial location and the amplitude of the minimum $\delta v_{\rm rot}$ of the rotation curve of the PDS~70 disk is best matched by the perturbations created by a 10~\Mjup \ planet. 
We note, however, that our hydrodynamic models consider a vertically isothermal temperature structure, whereas in a more realistic approach (introducing a more physical prescription for the vertical temperature structure), the deviation from Keplerian rotation may be higher in the disk surface than in the midplane, implying that the $\delta v_{\rm rot}$ in the current models may be underestimated \citep[Bae et al in prep.; see also Fig. 3 of][]{Teague18}. Relaxing the isothermal assumption and introducing a more physical prescription of the vertical temperature structure may be able to solve this discrepancy, but is beyond the scope of this study.

Toward the inner region the observed rotation curve is flatter, which may again be due to a slightly different gap shape (i.e., a flatter inner edge). The most conspicuous difference to the models is the region of super-Keplerian rotation beyond the planet, which extend farther out than in the models.  
As shown in Sect. \ref{sect:rotation_curve}, within the uncertainties, the observed rotation curve returns to Keplerian rotation close to the location of maximum emission in the continuum ring ($\sim$0.65\arcsec \ or 74 au) (see Fig. \ref{fig:rotation_curve}). This is consistent with the interpretation of large grains being trapped in the region of maximum pressure \citep[e.g., ][]{Pinilla12}. 
While we have shown that the observed integrated flux density profile can be reproduced well by one planet of 5 \Mjup, the large extension of super-Keplerian rotation and the concomitant far-out location of the continuum ring imply that the gap is wider in reality than predicted by all the models.
It therefore appears within our model assumptions that only one planet located at the orbit of PDS\,70\,b may not be sufficient to generate a kinematic signature in the disk with the inferred width or maintain a continuum ring at $\sim$74 au. This scenario needs to be probed by future observations at higher spectral resolution.

This is consistent with gap width considerations in the literature. As an example, hydrodynamical and dust simulations suggest that the accumulation of large dust grains is expected to be found at roughly 10 $\rm R_{H}$ outward of the planetary orbit \citep{Pinilla12,Rosotti+16}. For a 10~\Mjup \ planet at the location of the PDS\,70\,b orbit, the dust ring would therefore be expected at about 46-56 au, assuming a stellar mass of 0.88\,\Msun. 

This suggests that an additional low-mass planet located beyond PDS\,70\, or the combination with other physical mechanisms such as photoevaporation or dead zones may be needed to explain the outward-shifted location of the pressure bump. Models indeed predict that large gaps in transitional disks can be reproduced by introducing multiple planets \citep{Dodson-Robinson+11,Zhu+11,2015ApJ...802...42D}. Detailed modeling of the system by introducing multiple planets as well as deep observations are required to constrain the planetary architecture that is responsible for the observed features; this is beyond the scope of this study. 

An alternative scenario to explain the distant location of the ring compared to the position of PDS\,70\,b is to consider that the ring traces a secondary pressure bump. Single planets can indeed open multiple gaps in a disk with low viscosity, or alternatively, vortices generated at the edge of a gap can lead to a secondary ring \citep{Gomes+15}. In this latter scenario, the primary ring, located at $\sim$50 au (corresponding to $\sim$10 $\rm R_H$ from a 5 \Mjup \ planet at 22 au), would be depleted. The secondary ring would be located at $\sim$1.5$\times$ the location of the primary ring \citep{Gomes+15}, corresponding to $\sim$75 au, which is where the dust ring is found in the PDS\,70 system. Furthermore, secondary vortices may be generated at the edge of the secondary ring. If this is the case, this may also explain the azimuthal asymmetry observed in the dust continuum. A detailed exploration of this scenario will be the subject of a follow-up study.

\subsection{Upper limit on CPD dust mass}
The detection of H$\alpha$ emission at the location PDS\,70 implies that PDS\,70\,b is actively accreting \citep{Wagner2018} and therefore likely possesses an accretion disk. 
Still, we can only derive upper limits on the circumplanetary disk with our data. 
Models of planet formation predict circumplanetary dust around young planets, implying that CPDs should be frequent. 
However, searches for circumplanetary material in the submillimeter/millimeter continuum around other young substellar companions have been unsuccessful, although active accretion through the H$\alpha$ and/or Pa$\beta$ lines was detected in some of these cases \citep[e.g.,][]{Isella+14,Bowler+15,MacGregor+16,Wolff+17,Ricci+17,Pineda18}. Our upper limit on the CPD dust content of $\sim$0.01 \Mearth \ is similar to that derived for other systems \citep{Pineda18}. 

The detection of CPDs in the (sub-)millimeter regime may be challenging for several reasons.  First, CPDs are expected to be very small, which substantially reduces the emitting area and therefore the expected signal. 
Second, because the large grains are substantially trapped in the outer dust ring, the replenishment of large grains within the gap is expected to be inefficient. 
Even if small grains pass the gap and replenish the CPD, the radial drift is expected to be extremely efficient when they grow. The radial drift will deplete the large grains very fast \citep{Pinilla+13,zhu18}. 
A search for the CPD using gas kinematics as a tracer or NIR observations might therefore be more promising.

\section{Summary and conclusions}\label{sec:conclusions}
The young planet PDS\,70\,b is the most robust case of a directly imaged forming planet in the gap of a transition disk. We obtained ALMA Band 7 DDT observations in Cycle 5 and combined them with previous Cycle 3 data \citep{Long18} to study the natal environment of the planet at high angular resolution ($\sim$0.07\arcsec) in dust continuum and at the  $^{12}$CO J=3-2 transition. Our conclusions are listed below. 

\begin{itemize}
\item We detected the emission from the dust continuum as a highly structured ring. Its radial distribution peaks at $\sim$74 au. The inner edge of the ring shows evidence of a marginally resolved second ring component that peaks at around 60 au. We also detected a spur that projects into the gap at a PA of about 285$^{o}$ and confirmed an azimuthal brightness asymmetry with a brightness enhancement of about 13$\%$ in the northwest part of the ring.
\item We derived upper limits on the circumplanetary disk. Based on the noise level of the image we infer a 5$\sigma$ upper dust limit lower than $\sim$0.01 \Mearth.

\item The CO-integrated intensity shows evidence of two radial intensity depressions; the inner depression of the flux density lies at $\sim$0.2$\arcsec$ (corresponding to the location of PDS\,70\,b) and a second gap at about 0.6$\arcsec$. The inner gap is most likely carved by PDS\,70\,b. Comparison of the flux density profile to hydrodynamical simulations showed that the gap width and depth is best reproduced by a 5 \Mjup \ body. The outer gap can be explained by the dust being optically thick. Furthermore, we found evidence for an azimuthal intensity modulation that is due to self-absorption by optically thick CO. We also detected a bridge-like feature in the CO at the location of the spur seen in the continuum as well as the inner disk, which extended out to $\sim$15 au. Finally, we reported the tentative detection of a possible point source in the $^{12}$CO emission maps, the existence of which needs to be confirmed with additional observations.

\item We detected significant deviation from Keplerian rotation inside $\sim$0.8\arcsec. The width of the $\delta v_{\rm rot}$ feature is consistent with the far-out location of the dust ring. Comparison to hydrodynamical simulations implies that the depth of the kinematic signature is best matched by a $\sim$10 \Mjup \ object (within our model assumptions of an isothermal disk), but the width of the feature suggests that one planet alone located at the orbit of PDS\,70\,b may not be sufficient to generate a gap with the inferred extension. 
An additional physical mechanisms or a second low-mass body may be required to explain the disk morphology. Future observations at higher angular and spectral resolution will allow us to place tighter constraints on the planetary system architecture that can account for all of the observed features in the PDS\,70 disk morphology. 
\end{itemize}

\begin{acknowledgements}
We thank the anonymous referee for constructive comments that helped improve the manuscript. 
This paper makes use of the following ALMA data: ADS/JAO.ALMA$\#$2017.A.00006.S, ADS/JAO.ALMA$\#$2015.1.00888.S.
ALMA is a partnership of ESO (representing its member states), NSF (USA) and NINS (Japan), together with NRC (Canada) and NSC and ASIAA (Taiwan) and KASI (Republic of Korea), in cooperation with the Republic of Chile. The Joint ALMA Observatory is operated by ESO, AUI/NRAO and NAOJ.
The National Radio Astronomy Observatory is a facility of the National Science Foundation operated under cooperative agreement by Associated Universities, Inc.
This publication has received funding from the European Union’s Horizon 2020 research and innovation program under grant agreement No 730562 (RadioNet). %M.K.
R.T. acknowledges funding from the NSF grants AST-1514670 and NASA NNX16AB48G.
M.B. acknowledges funding from ANR of France under contract number ANR-16-CE31-0013 (Planet Forming disks).
J.B. acknowledges support from NASA grant NNX17AE31G and computing resources provided by the Advanced Research Computing at the University of Michigan, Ann Arbor.
Y.L. acknowledges supports by the Natural Science Foundation of Jiangsu Province of China (Grant No. BK20181513) and by the Natural Science Foundation of China (Grant No. 11503087).
S.F. acknowledges an ESO Fellowship.
G.R. and A.J. are supported by the DISCSIM project, grant agreement 341137 funded by the European Research Council under ERC-2013-ADG.
G.R. acknowledges support from the Netherlands Organisation for Scientific Research (NWO, program number 016.Veni.192.233). 
GHMB and M.F. acknowledge funding from the European Research Council (ERC) under the European Union’s Horizon 2020 research and innovation programme (grant agreement No. 757957).
L.P. acknowledges support from CONICYT project Basal AFB-170002 and from FONDECYT Iniciaci\'{o}n project No. 11181068. 
A.M. acknowledges the support of the DFG priority program SPP 1992 "Exploring the Diversity of Extrasolar Planets" (MU 4172/1-1).
\end{acknowledgements}

\bibliography{bibliography}
\bibliographystyle{aa} % style aa.bst

\appendix
\counterwithin{figure}{section}

\section{Appendix information}

\subsection{Dependency of CPD detection limits on planetary mass and accretion rate}\label{app:CPD}
We explore the dependency of the CPD detection limit on the planetary mass and accretion rate. 
In our approach, the CPD temperature profile results from three contributions: heating from irradiation by the central star, from irradiation by the planet, and from viscous accretion (see Equ. \ref{equ:CPD_temp}). The irradiation by the planet depends on its luminosity, and therefore on its temperature and radius \citep[$T_P\sim1200$\,K, $r_P\sim$3 \Rjup \ for PDS\,70\,b; ][]{mueller2018}, and the contribution from accretional heating is proportional to the product of planet mass and accretion rate (see Equ. \ref{equ:irr} and \ref{equ:acc}). 
Their relative contribution can be expressed as

\begin{equation}
    \frac{T_{acc}^4}{ T_{irr,p}^4} (r) = \frac{30 G}{8 \pi \sigma_{SB} T_p^4 r_p^2} \frac{M_p \dot{M}_{acc}}{r}
    \left[ 1 - \left( \frac{r_p}{r} \right)^{1/2} \right]
\end{equation}
This expression has a maximum at about 2.25 $\times$ $r_p$, and we can therefore write

\begin{equation}
\begin{split}
    \frac{T_{acc}^4}{T_{irr,p}^4} &\leq \frac{5 G M_p \dot{M}_{acc}}{18 \pi \sigma_{SB} T_p^4 r_p^3} \\
        &\approx 0.03 \left(\frac{1200 \rm K}{T_P} \right)^4 \times \left(\frac{3 \ \rm R_{Jup}}{R_P} \right)^3 \times \left( \frac{M_P}{5\ \rm M_{Jup}} \right) \times \left(\frac{\dot{M}_{acc}}{10^{-8} \ \rm M_{Jup}/yr} \right)
\end{split}
\end{equation}
For the given parameter choice, accretional heating is therefore negligible. 

For a planetary mass of 5 \Mjup \ and radius of 3 \Rjup, \cite{Wagner2018} calculated an accretion rate for PDS\,70\,b of 3$\times10^{-9}$ to 9$\times 10^{-8}$ \Mjup/yr, depending on the assumed extinction. Thus, even in case of high extinction and therefore high accretion rate, the term $T_{acc}^{4}/T_{irr,p}^{4}$ is lower than 0.3, and the contribution from accretional heating is still marginal. We note that the planetary mass cannot be varied independently of the accretion rate because the product $M_P \dot{M}_{acc}$ is to be conserved. We therefore conclude that the temperature structure in our calculation is insensitive to the choice of the planet mass within the estimated range for a CPD of a given size.

\begin{figure}[hbt]
\centering
\includegraphics[width=0.5\textwidth]{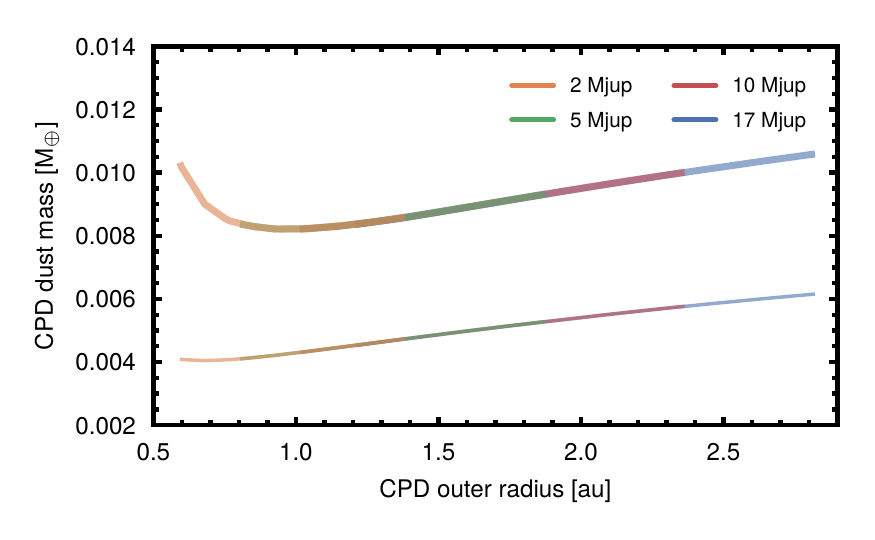}
\caption{CPD 5$\sigma$ (thick line) and 3-$\sigma$ (thin line) detection limits for different planet masses covering the mass range for PDS\,70\,b estimated by \cite{keppler18} and \cite{mueller2018}. \label{fig:CPD_limits_lum}}
\end{figure}

The outer outer radius does depend on the planet mass, however. The lower the planet mass, the smaller the disk it will be able to retain. This is illustrated in Fig. \ref{fig:CPD_limits_lum}, which shows the 5$\sigma$ and 3$\sigma$ detection limits for the estimated mass ranges based on the comparison of NIR photometry with evolutionary models and atmospheric modeling \citep{keppler18,mueller2018}. 
The figure illustrates that 1) at a given CPD size, the CPD flux is independent of the planetary mass, 2) the 5$\sigma$ detection limits (thick lines) are rather constant for all disk sizes, mostly below $\sim$0.01\,\Mearth, indicating optically thin emission (except for the case of 2 \Mjup, where emission is in transition to optically thick), and 3) CPDs around planets with different masses cover different ranges of disk sizes. 

\subsection{Effect of beam smearing on the rotation curve}\label{app:beam_smearing}

For a beam of a given size, the spectrum, and thus the velocity of the gas traced, observed at each pixel corresponds to the average of all spectra within the beam centered on that pixel, weighted by their line intensities and the beam shape. If the intensity gradient across the beam is steep, this will cause the sampled velocity to be strongly biased toward the region of highest intensity, rather than the beam center. 

\begin{figure*}[hbt]
\centering
\includegraphics[]{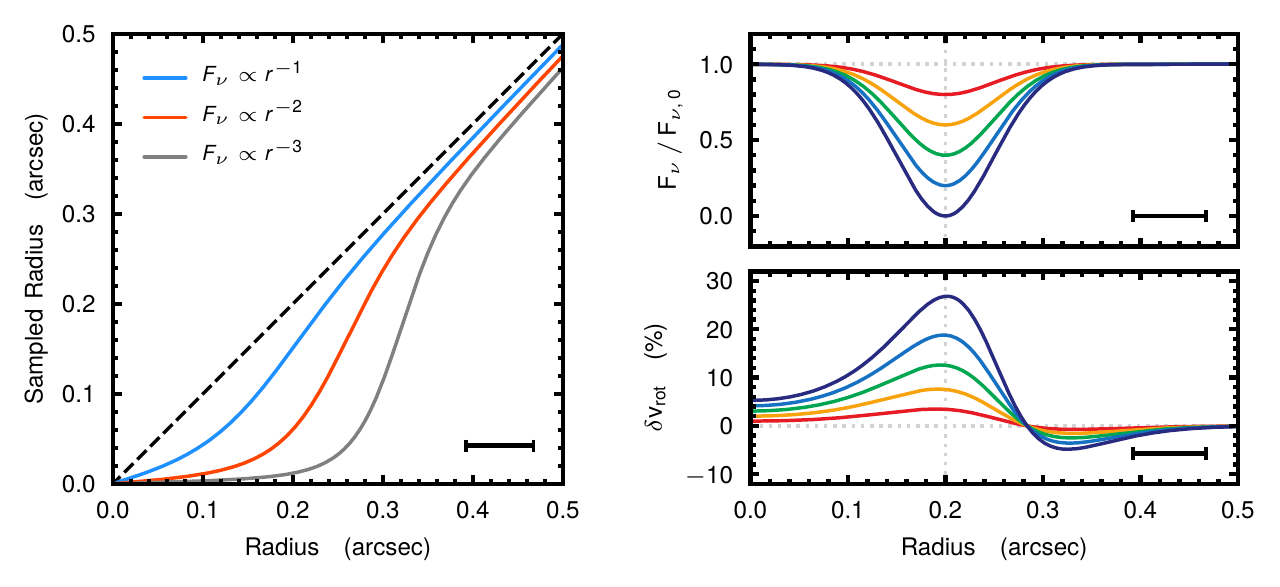}
\caption{Effect of beam convolution in the presence of intensity gradients on the radial sampling of the rotation velocity. 
The left panel illustrates the deviation of the sampled radius (i.e. the radius within the beam at which the intensity and therefore the weighting of the velocity is highest due to the convolution with the beam) from the real radius in the presence of smooth intensity profiles with different power-law indices. The effect is stronger for steeper intensity profiles. 
The right panels shows the effect of beam convolution in the presence of a gap-shaped intensity profile with varying depths (upper right) on the resulting residual rotation curve (bottom right). 
In both cases, a beam size of 76 mas is assumed, shown by the horizontal black bar. 
 \label{fig:beam_smear_analytical}}
\end{figure*}

This effect is illustrated in Fig. \ref{fig:beam_smear_analytical} (left). For a smooth radial disk intensity profile, the figure shows for each distance the sampled radius, that is, the radius within the beam at which the velocity receives the highest weighting and which therefore corresponds to the effective radius at which the velocity is observed. This is shown for different power-law exponents of the radial intensity profile of the disk. We note that the steeper the intensity profile, the greater the bias of the sampled radius toward smaller radii, and the greater the overestimate of the measured velocity.

This is even more complex when the intensity profile deviates from a simple power law, as in the presence of a gap structure. The additional steep gradients at the gap edges cause regions closer to the inner gap edge to become even more biased toward smaller distances and therefore higher velocities, whereas regions close to the outer gap edge are biased toward larger distances and lower velocities. Figure \ref{fig:beam_smear_analytical}  (bottom right) shows the deviation from Keplerian rotation, assuming an intensity profile with a gap structure centered around 0.2\arcsec \ and a beam size of 76 mas (top right). The resulting $\delta \rm{v_{rot}}$ profile is asymmetric with respect to the gap center, with super-Keplerian rotation in the inner regions changing into sub-Keplerian rotation beyond $\sim$ 0.3 \arcsec, and the strength of the deviation is sensitive to the gap depth. This beam-smearing effect is added to the deviation from Keplerian rotation that is due to the planet-induced pressure gradient. 

\begin{figure*}[hbt]
\centering
\includegraphics[]{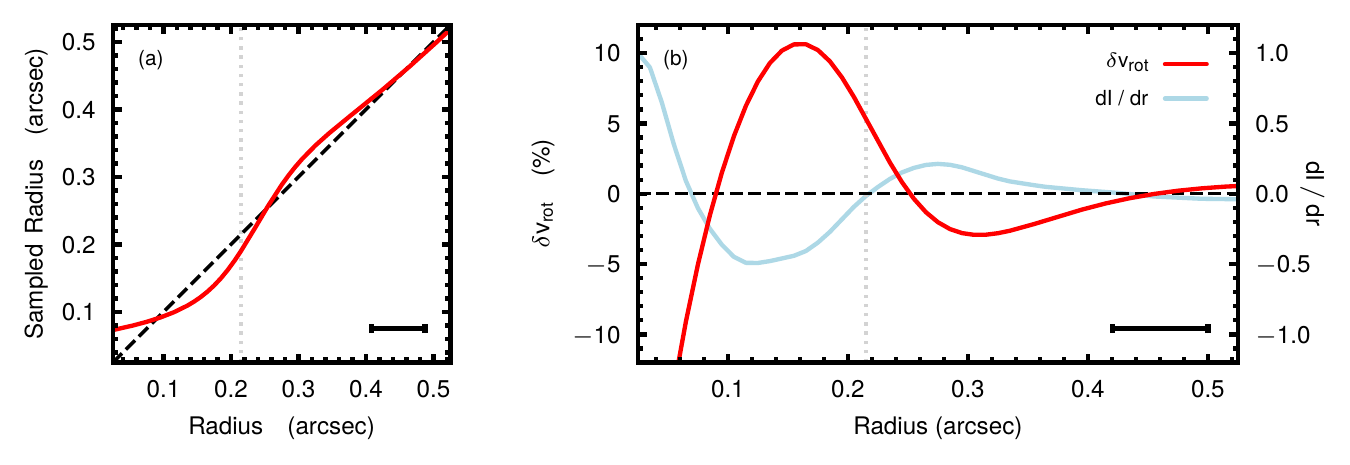}
\caption{Similar to Fig.~\ref{fig:beam_smear_analytical}, but using the observed intensity profile from Fig.~\ref{fig:zeroth_moment_profiles}. The left panel shows the bias in the sampled radius caused by the gap at $0.2\arcsec$, shown by the vertical dashed line. The right panel shows the resulting deviation in velocity expected by this bias in red. For contrast, the blue line shows the deviations expected from a pressure gradient, here using the normalized radial gradient of the intensity as a proxy. The recovered $\delta v_{\rm rot}$ therefore is a combination of these two effects. \label{fig:beam_smear_observations}}
\end{figure*}

Figure~\ref{fig:beam_smear_observations} demonstrates the effects of this bias using the radial intensity profile from Fig.~\ref{fig:zeroth_moment_profiles} and compares this effect to the functional form from the pressure gradient (shown in light blue). The resulting profile is the combination of both factors, whose relative amplitudes depend on the gap shape. While this limits interpretation, these effects are fully accounted for with forward modeling, as presented in Section~\ref{sec:modelling}.

\begin{figure*}[hbt]
\centering
\includegraphics[width=0.7\textwidth]{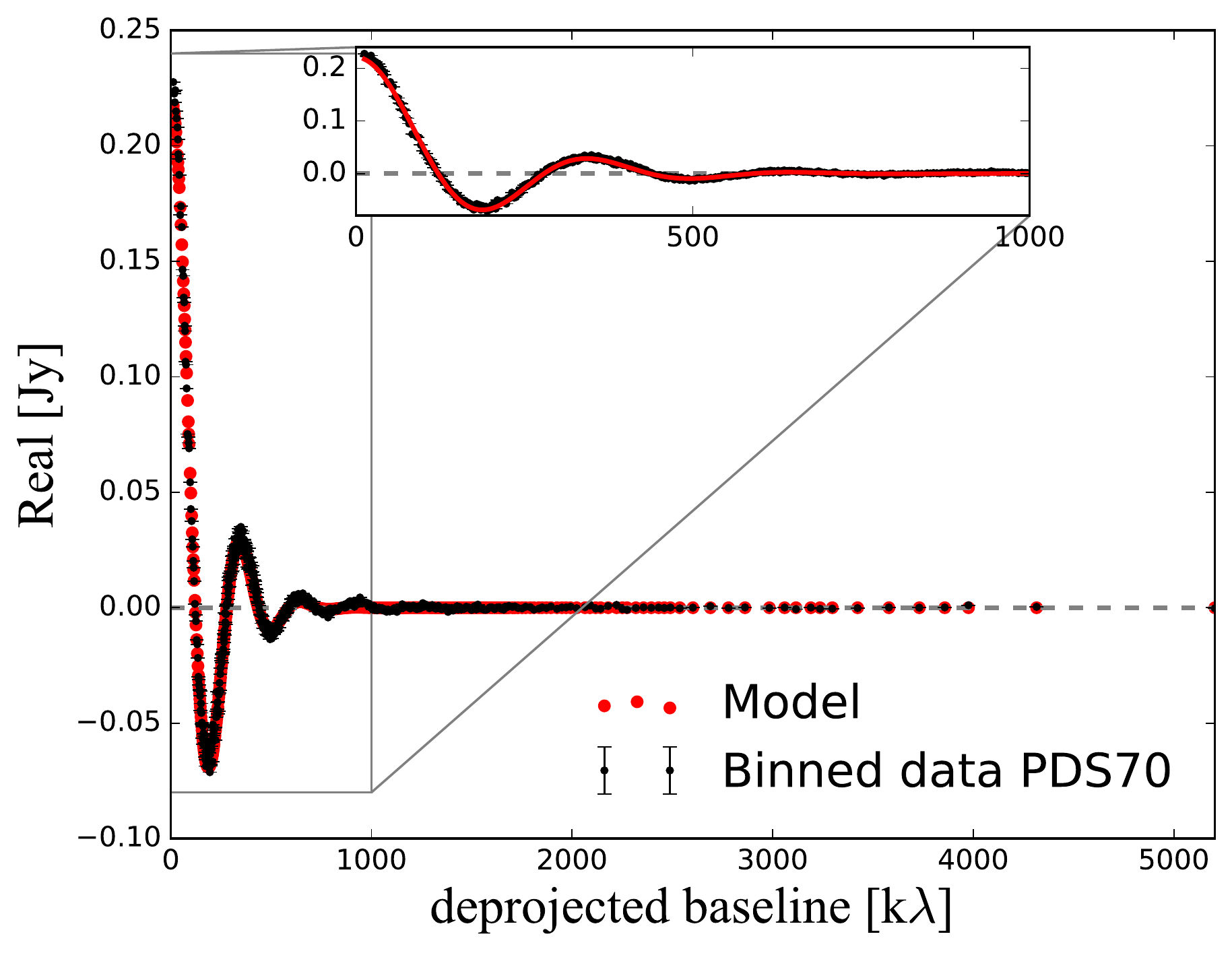}
\caption{Results on the MCMC fit of the deprojected and binned visibilities of the dust continuum, following the approach by \cite{2018ApJ...859...32P}.  \label{fig:MCMC_dust}}
\end{figure*}

\begin{figure*}[hbt]
\centering
\includegraphics[width=1.0\textwidth]{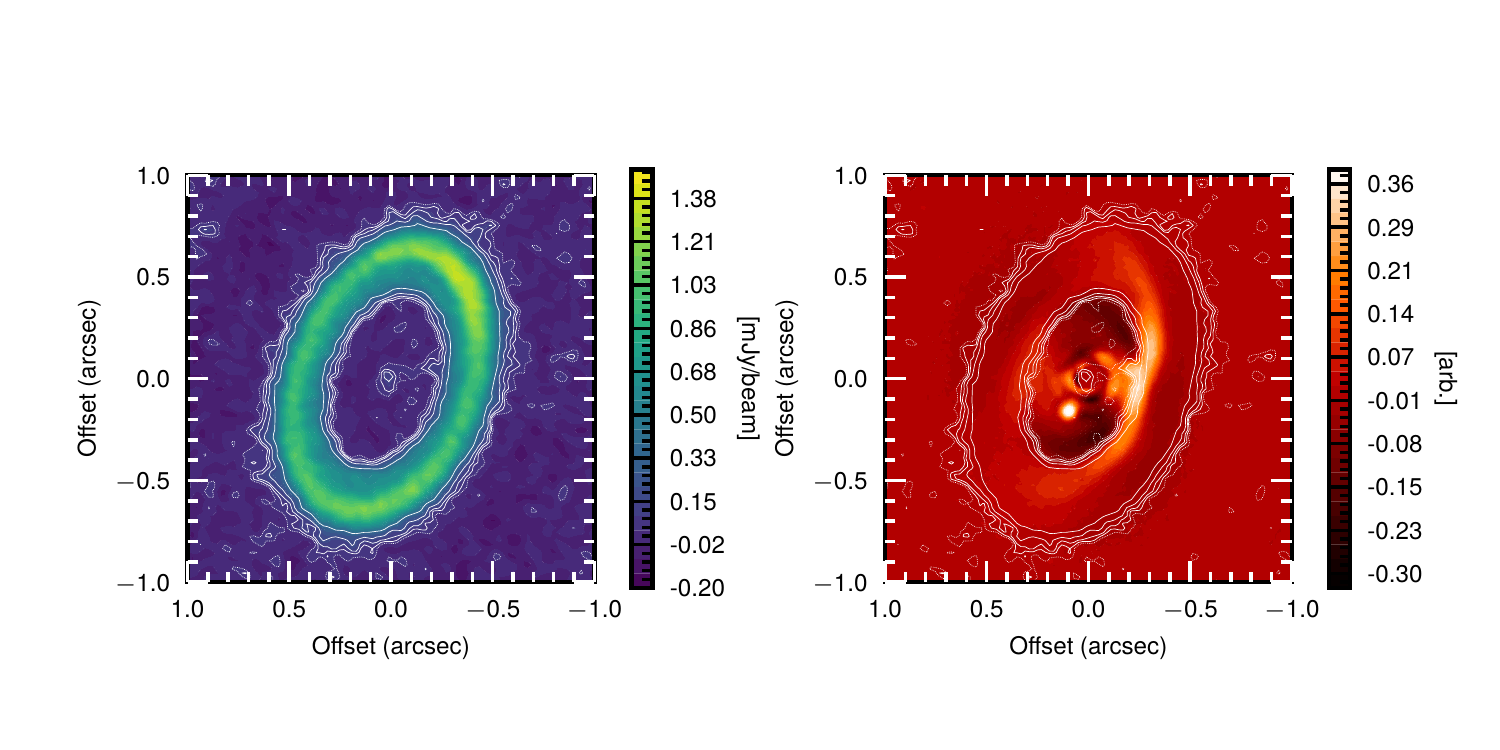}
\caption{
Cycle 5 dust continuum data (\textit{left}) and SPHERE NIR image \citep[\textit{right}, from ][]{mueller2018}, the ALMA Cycle 5 data contours are overlaid in white. The solid contours show levels of 2,3,5, and 10 $\sigma$, and the dotted line corresponds to 1 $\sigma$. 
} \label{fig:DDT_dust}
\end{figure*}

\end{document}